\documentclass[a4paper,12pt]{article}
\usepackage{jheppub} 
\usepackage{lineno}
\usepackage[english,activeacute]{babel}
\usepackage{mathtools}
\usepackage{xcolor}
\usepackage[scr=boondox,  
            cal=esstix]   
           {mathalpha}

\title{\boldmath Classical Kerr-Schild double copy in bigravity for maximally symmetric spacetimes}

 \author[1]{H. Garc\'{\i}a-Compe\'an\note{hugo.compean@cinvestav.mx}}
 \author[1]{and C. Ramos\note{cesar.ramos@cinvestav.mx}}
 \affiliation{Departamento de F\'{\i}sica,\\
 Centro de Investigaci\'on y de Estudios Avanzados del Instituto Polit\'ecnico Nacional,\\
 P.O. box 14-740, C.P. 07000, Ciudad de M\'exico, Mexico}

\abstract{A generalized Kerr-Schild ansatz for bigravity, already considered in the literature, which leads to linear interactions between the metrics is used to study the bigravity equations in the context of the double copy. By contracting the resulting spin-2 field bigravity equations of motion using Killing vector fields, as is usually carried out in general relativity, we arrive to the single and zeroth copy equations for the mentioned ansatz. For the case of stationary solutions, we obtain two Maxwell and two conformally coupled scalar field equations for the single and zeroth copies respectively, and the linear interactions are absent. In the time-dependent case we obtain equations for the fields which are coupled. By decoupling these equations and at the zeroth copy level, we recover a massless and a massive field whose mass is proportional to the Fierz-Pauli mass and depends on the coefficients of the interaction potential between the metrics. This has been also previously documented in the literature and is now reinterpreted within the context of the double copy proposal.}

\begin{document}
\maketitle
\flushbottom
\section{Introduction} 
In the search of a quantum theory of gravity, many developments have been considered in order to find some viewpoints that allows
to consider new approaches that relate gravity and field theories
\cite{Gupta:1954zz,Weinberg:1965rz,Weinberg:1980kq, Wald:1986bj,Feynman:1996kb}. One of these approaches is the one of the double copy (for some recent reviews, see \cite{Bern:2019prr,Bern:2022wqg,Adamo:2022dcm}). This proposal was first discussed in the context of scattering amplitudes of gravity and gauge theories \cite{Bern:2008qj,Bern:2010yg,Bern:2010ue} and further extended to a relation between the solutions of the classical equations of motion of both theories, this is termed the classical double copy \cite{Monteiro:2014cda,Luna:2015paa,Ridgway:2015fdl,Luna:2016due,Bahjat-Abbas:2017htu,Carrillo-Gonzalez:2017iyj,Bern:2019prr,Alkac:2021bav}. This double copy is realized at the level of the classical equations of motion of the gauge and gravity theories. Many developments have been described at the linear level but some other articles have discussed the case of non-linear extensions. The classical double copy has been proposed for spacetimes of the Kerr-Schild form \cite{Kerr:1965wfc} and double Kerr-Schild form \cite{Luna:2015paa}. The double copy can be carried out in another representation called the Weyl double copy
\cite{Luna:2018dpt,Sabharwal:2019ngs,Alawadhi:2020jrv,Godazgar:2020zbv}, where a proposition to address sources was considered in \cite{Easson:2021asd} and further progress in \cite{Easson:2022zoh}. Many interesting results have been achieved using this approach, which has been shown to be equivalent to the Kerr-Schild one. Moreover, there is a generalization to the Weyl double copy which is the twistor double copy \cite{White:2020sfn,Chacon:2021wbr,Chacon:2021hfe,Chacon:2021lox,Luna:2022dxo}. Interesting results can be obtained in this context. In the present article we will work in the context of the Kerr-Schild type metrics.  

On the other hand, the theories of massive gravity are very important in order to study the possibility for the existence of mass for the graviton avoiding the inconsistencies that arise in the process of reaching this goal (for a review, see \cite{deRham:2014zqa}). Among some different proposals in various dimensions, the bigravity theory is one of the important proposals 
\cite{Hassan:2011zd,Hassan:2011vm} and further explored in different contexts in 
\cite{Banados:2011hk,Deffayet:2011uk,Pilo:2011zz,Comelli:2011wq,Volkov:2012wp,Comelli:2014bqa,Comelli:2015pua,Ayon-Beato:2015qtt,Ayon-Beato:2018hxz,Lust:2021jps}. In the present article we study the bigravity in the context of the Kerr-Schild double copy. There are in the literature some descriptions of the massive gravity in the context of the double copy, see for instance, \cite{Momeni:2020vvr,Johnson:2020pny,Gonzalez:2021bes,CarrilloGonzalez:2022mxx,Lust:2023sfk}, but bigravity has not been extensively studied in the context of the double copy.

The present paper is organized as follows: In Sec. \ref{prelim} we shortly review the double copy procedure and give some relevant definitions of bigravity. In Sec. \ref{DKS_ansatz} we consider a general Kerr-Schild ansatz  but we restrict ourselves to maximally symmetric spacetimes in bigravity and linear interactions between the metrics. By rewriting the Ricci tensor for the metrics and contracting with Killing vectors we obtain equations for spin-1 and spin-0 which we study in Sec. \ref{BG_DC}, for stationary and time-dependent solutions, including some examples and in Sec. \ref{final} we conclude with some final comments.

\section{Preliminaries}\label{prelim}
In this section we comment on the double copy by introducing the Kerr-Schild double copy in general relativity. Then, we briefly overview bigravity from a perspective that will be useful for future reference.

\subsection{Classical Kerr-Schild double copy}
The double copy is a mathematical relationship discovered through the study of scattering amplitudes in both gravity and gauge theories \cite{Bern:2008qj, Bern:2010yg, Bern:2010ue}, rooted in a duality between color and kinematic factors that appear in the amplitudes, known as the BCJ duality. The Kerr-Schild classical double copy is an analogous procedure hypothesized to exist within quantum field theory \cite{ Bern:2010yg, Bern:2010ue}, which considers that certain classical gauge and gravity solutions can be related by a double copy relation and involves the Kerr-Schild metrics \cite{Kerr:1965wfc, Gurses:1975vu, Stephani:2003tm}, a special family of solutions of the Einstein equations in general relativity.

In the case of general relativity, this classical approach was studied in \cite{Monteiro:2014cda} for Minkowski background and in maximally symmetric spacetimes in \cite{Carrillo-Gonzalez:2017iyj}. The Kerr-Schild ansatz in curved backgrounds has the following form:
\begin{align}
    \label{ks_1}
    g_{\mu\nu}= \overline{g}_{\mu\nu}+\kappa h_{\mu\nu} \, ,  \quad h_{\mu\nu} = \phi k_{\mu}k_{\nu} \, , 
\end{align}
where $\kappa^{2}=16\pi G_{N}$ is the gravitational coupling, $G_{N}$ is the Newton's constant and $\overline{g}_{\mu\nu}$ is the background metric. We have also defined the perturbation to the background metric, $h_{\mu\nu}$, where  $\phi$ is a scalar function and $k_{\mu}$ is a vector which is null with respect to the two metrics and geodesic with respect to $g_{\mu\nu}$, so it is also geodesic with respect to $\overline{g}_{\mu\nu}$ \cite{Stephani:2003tm}, so that the following holds:
\begin{align}
    \nonumber g^{\mu\nu} k_{\mu} k_{\nu} = 0  =  \overline{g}^{\mu\nu} k_{\mu} k_{\nu} \, , \quad k^{\mu} \nabla_{\mu} k_{\nu} = 0 =  k^{\mu} \overline{\nabla}_{\mu} k_{\nu} \, . 
\end{align}
The covariant derivative $\nabla_{\mu}$ is associated to the metric $g_{\mu\nu}$ by the compatibility condition $\nabla_{\lambda}\, g_{\mu\nu}=0 \,$ and  $\overline{\nabla}_{\mu}$ is the covariant derivative associated to $\overline{g}_{\mu\nu}$ by the condition $\overline{\nabla}_{\lambda}\, \overline{g}_{\mu\nu}=0$. This permit us to write
\begin{align}
    \label{ric_1}
    R^{\mu}{}_{\nu}=\overline{R}^{\mu}{}_{\nu}-\kappa\, h^{\mu\lambda} \overline{R}_{\lambda \nu}+\frac{\kappa}{2}\overline{\nabla}_{\lambda}\left[\overline{\nabla}^{\mu}h^{\lambda}{}_{\nu}+\overline{\nabla}_{\nu}h^{\mu\lambda}-\overline{\nabla}^{\lambda}h^{\mu}{}_{\nu}\right] \ .
\end{align}
For the ansatz (\ref{ks_1}), this is the specific combination of the indices which linearizes the Ricci tensor, with $R^{\mu}{}_{\nu}=g^{\alpha\mu}R_{\alpha\nu}$ and $\overline{\nabla}^{\mu}=\overline{g}^{\mu\alpha}\overline{\nabla}_{\alpha}$. The form of (\ref{ric_1}) is what allows us to elucidate which equations the gauge fields will satisfy, as can be seen in \cite{Monteiro:2014cda, Carrillo-Gonzalez:2017iyj}. 

The classical Kerr-Schild double copy procedure permit us to relate gauge and gravity classical solutions, so that if we have a solution in the gravity side, it corresponds to an associated solution in the gauge side via this procedure. The Kerr-Schild ansatz (\ref{ks_1}) possesses the property that the graviton factorizes into an outer product of two Lorentz vectors accompanied by a scalar function. To get from the double copy field, which is $h_{\mu\nu}$, to the single copy field, we need to remove one Lorentz factor $k_{\mu}$ and replace it with a color factor $c^{a}$ to obtain the non-abelian single copy field $A^{a}_{\mu}$,
\begin{align}
    \nonumber A^{a}_{\mu}=c^{a}\, \phi\,k_{\mu}\, , \quad \frac{\kappa}{2} \rightarrow g \, ,
\end{align}
and we also have to map the sources and coupling constant from the gravity theory to the sources and constants from the gauge theory. Similarly, for the zeroth copy, we eliminate another $k_{\mu}$ and add a color factor $c^{a'}$, and also replace the sources and constants, 
\begin{align}
     \nonumber \Phi^{aa'}=c^{a}c^{a'} \phi\, , \quad g \rightarrow y  \,,
\end{align}
and obtain the biadjoint scalar field $\Phi^{aa'}$. By considering stationary solutions and a flat background for the ansatz (\ref{ks_1}), $\overline{g}_{\mu\nu}=\eta_{\mu\nu}$, in \cite{Monteiro:2014cda}, it was obtained that, after stripping the color factors, the gauge field $A_{\mu} = \phi k_{\mu}$ solves:
\begin{align}
    \nonumber \partial_{\mu}F^{\mu\nu}=0 \, .
\end{align}
The field $A_{\mu}$ is the single copy of $h_{\mu\nu}$, which is obtained by removing one Lorentz factor $k_{\nu}$ from the graviton $h_{\mu\nu}$. By removing the Lorentz factor in $A_{\mu}$, i.e., taking the zeroth copy, we obtain the field $\phi$, which satisfies
\begin{align}
    \nabla^{2}\phi =0\, .
\end{align}
These are the abelian version of the fields. For the present work, we aim to explore the possibility to extend this type of analysis to a different theory of gravity, specifically considering a massive theory of gravity, such as bimetric gravity. 

\subsection{Theory of massive bigravity}
From the perspective of particle physics, we can consider that the gravitational field is mediated by a massless spin-2 particle, known as the graviton. In fact, under certain assumptions such as locality, Lorentz invariance and unitarity, general relativity is the unique theory describing a single interacting massless spin-2 particle which couples to matter and energy \cite{Gupta:1954zz,Weinberg:1965rz,Weinberg:1980kq, Wald:1986bj,Feynman:1996kb}. In order to find an alternative to Einstein's theory of gravity, one can modify the assumption of the vanishing rest-mass of the graviton and consider the particle to be massive. In this context is where we have bigravity theory, which is a massive theory of gravity.

Bigravity (or bimetric gravity) is a ghost free non-linear massive gravity theory in four dimensions proposed by Hassan and Rosen \cite{Hassan:2011zd}, which considers two dynamical metrics $g_{\mu\nu}$ and $f_{\mu\nu}$ that are interacting with each other. The interaction between the metrics is described by a non-derivative interaction potential $\mathcal{U}[g,f]$. The theory propagates $5+2$ degrees of freedom, which results from the fact that we can consider bigravity as a theory of one massive and one massless spin-2 field. The action for bigravity can be read as:
{\small\begin{align}
    \label{bigrav_action}
    S_{bi}[g, f]= & \frac{{M_{g}}^2}{2} \int d^{4} x \sqrt{-g} \, R[g]+\frac{{M_{f}}^2}{2} \int d^{4} x \sqrt{-f} \mathcal{R}[f] -m^{2} {M_{\text{eff}}}^2 \int d^{4} x \sqrt{-g} \, \mathcal{U}[g, f],
\end{align}}where $M_{g}$ and $M_{f}$ are the Planck masses for each of the metrics $g_{\mu\nu}$ and $f_{\mu\nu}$ respectively and
\begin{align*}
    \nonumber {M_{\text{eff}}}^2 = \frac{(M_{g}M_{f})^2}{M_{g}^{2}+M_{f}^2} \, 
\end{align*} 
is the ``effective'' Planck mass. These Planck masses are related to the gravitational constants for each metric by $8\pi G_{g}={M_g}^{-2}$ and $8\pi G_{f}={M_{f}}^{-2}$ in natural units. The action can be rewritten in terms of the gravitational couplings $\kappa_{g}$ and $\kappa_{f}$  for each of the metrics, i.e. ${\kappa_{g}}^{2}=16\pi G_{g}$ and ${\kappa_{f}}^{2}=16\pi G_{f}$ so ${\kappa_{g}}^2=2 {M_{g}}^{-2}$ and ${\kappa_{f}}^{2}=2 {M_{f}}^{-2}$. The interaction potential is written in terms of the matrix $\gamma^{\mu}{}_{\nu}$, which encodes the interaction between the metrics
\begin{align}
    \label{gamma_def}
    \mathcal{U}[g,f] & =\sum_{k=0}^{4} b_{k} \mathcal{U}_{k}(\gamma) \, , \qquad \gamma^{\mu}{}_{\nu}=\sqrt{g^{\mu\alpha}f_{\nu\alpha}} \, ,
\end{align}
where $b_{k}$ are coupling constants and we refer to $\gamma^{\mu}{}_{\nu}$ as the interaction matrix. 
The functions $\mathcal{U}_{k}(\gamma)$ are defined by 
\begin{align}
    \nonumber \mathcal{U}_{0}(\gamma)&=1\, ,  \quad 
    \mathcal{U}_{1}(\gamma)=[\gamma]\, , \quad 
    \mathcal{U}_{2}(\gamma)=\frac{1}{2 !}\left([\gamma]^{2}-[\gamma^{2}]\right)\,, \\
    \nonumber \mathcal{U}_{3}(\gamma)&=\frac{1}{3 !}\left([\gamma]^{3}-3[\gamma][\gamma^{2}]+2[\gamma^{3}]\right)\, , \\
    \nonumber \mathcal{U}_{4}(\gamma)&=\frac{1}{4 !}\left([\gamma]^{4}-6[\gamma]^{2}[\gamma^{2}]+8[\gamma][\gamma^{3}]+3[\gamma^{2}]^{2}-6[\gamma^{4}]\right) \, , 
\end{align}
where $(\gamma^{n})^{\mu}{}_{\nu}=\gamma^{\mu}{}_{\alpha_{1}}\gamma^{\alpha_{1}}{}_{\alpha_{2}}\cdots \gamma^{\alpha_{n-1}}{}_{\nu}$ and the trace of the matrix $\gamma^{\mu}{}_{\nu}$ is defined as $[\gamma]\equiv \operatorname{tr}(\gamma^{\mu}{}_{\nu})=\gamma^{\mu}{}_{\mu}\,$, which shows that the potential depends only on the metrics and not on their derivatives.  In the formulation of bigravity which uses a $\mathcal{K}^{\mu}{}_{\nu}$ matrix instead of $\mathcal{\gamma}^{\mu}{}_{\nu}$ and another set of constants $c_{k}$ instead of $b_{k}$, setting $c_{1}=1$ is done in order to obtain the Fierz-Pauli mass term, which traduces to $b_{2} =-1-2b_{3}-b_{4}$ in our case. This means that the action in (\ref{bigrav_action}) reduces to the Fierz-Pauli action at a certain limit \cite{Hassan:2011vm}. The variation of the action in (\ref{bigrav_action}) lead us to the equations of motion for bigravity, which can written in terms of $\kappa_{g}$, $\kappa_{f}$ and $\kappa$, a function of the gravitational couplings for each metric, so that we have:
\begin{align}
\label{bigrav_eom}
    G^{\mu}{ }_{\nu}=\frac{m^{2} {\kappa_{g}}^{2}}{{\kappa}^{2}} V^{\mu}{ }_{\nu} \,, \quad \ \ \ \ \ \ \mathcal{G}^{\mu}{}_{\nu}=\frac{m^{2} {{\kappa_{f}}^{2}}}{{\kappa}^{2}} \mathcal{V}^{\mu}{ }_{\nu} \, ,
\end{align}
where $\kappa^{2} \equiv {\kappa_{g}}^{2}+{\kappa_{f}}^{2}$ and $G^{\mu}{}_{\nu}$ and $\mathcal{G}^{\mu}{}_{\nu}$ are the Einstein tensors for the metrics $g_{\mu\nu}$ and $f_{\mu\nu}$ respectively. $V^{\mu}{}_{\nu}$ and $\mathcal{V}^{\mu}{}_{\nu}$ are defined as:
\begin{align}
    \nonumber V^{\mu}{}_{\nu} \equiv \frac{2 g^{\mu \alpha}}{\sqrt{-g}} \frac{\delta(\sqrt{-g} \, \mathcal{U})}{\delta g^{\alpha \nu}}=\tau^{\mu}{ }_{v}-\mathcal{U} \delta^{\mu}{ }_{\nu} \, , \quad \mathcal{V}^{\mu}{}_{\nu}  \equiv \frac{2 f^{\mu \alpha}}{\sqrt{-f}} \frac{\delta(\sqrt{-g} \, \mathcal{U})}{\delta f^{\alpha \nu}}=-\frac{\sqrt{-g}}{\sqrt{-f}} \tau^{\mu}{ }_{\nu},
\end{align}
which are the energy-momentum tensors arising from the interaction between the two metrics, where
\begin{align}
    \nonumber \tau^{\mu}{}_{\nu}= & \left(b_{1} \mathcal{U}_{0}+b_{2} \mathcal{U}_{1}+b_{3} \mathcal{U}_{2}+b_{4} \mathcal{U}_{3}\right) \gamma^{\mu}{ }_{\nu} -\left(b_{2} \mathcal{U}_{0}+b_{3} \mathcal{U}_{1}+b_{4} \mathcal{U}_{2}\right)(\gamma^{2})^{\mu}{ }_{\nu} \\
    \nonumber & +\left(b_{3} \mathcal{U}_{0}+b_{4} \mathcal{U}_{1}\right)(\gamma^{3})^{\mu}{ }_{\nu} -b_{4} \mathcal{U}_{0}(\gamma^{4})^{\mu}{ }_{\nu} \, . 
\end{align}
For simplicity, we define what we refer to as the ``interaction tensors'':
\begin{align}
    Q^{\mu}{}_{\nu} \equiv \frac{m^{2} {\kappa_{g}}^{2}}{{\kappa}^{2}}V^{\mu}{}_{\nu} \, , \quad  \ \ \ \ \ \mathcal{Q}^{\mu}{}_{\nu} \equiv \frac{m^{2} {\kappa_{f}}^{2}}{{\kappa}^{2}}\mathcal{V}^{\mu}{ }_{\nu} \, . 
\end{align}
We stick to the convention that the standard letters such $G_{\mu\nu}$ refer to quantities of the $g_{\mu\nu}$ metric and the cursive letters like $\mathcal{G}_{\mu\nu}$ refer to quantities of the $f_{\mu\nu}$ metric. We use the ``mostly plus'' signature convention $(-,+,+,+)$ through this text.

\section{Kerr-Schild ansatz in bigravity}\label{DKS_ansatz}

The strategy we will follow in this study involves to consider a generalization of the Kerr-Schild ansatz, where the background spacetimes are not restricted to be flat. From the perspective of the bimetric gravity theory, to consider non-flat spacetimes makes sense because the cosmological constant contribution for each of the metrics are considered in the definition of the interaction potential $\mathcal{U}[g,f]$. We advance our analysis in alignment with the approach outlined in \cite{Stephani:2003tm}, focusing on bigravity and curved backgrounds.

We start with the following generalized Kerr-Schild ansatz:
\begin{align}
\label{type_b_ansatz_1}
    \nonumber g_{\mu\nu}= &\overline{g}_{\mu\nu}+\kappa_{g}h_{\mu\nu} \,,  \quad  \ \ \ \ h_{\mu\nu} = \phi_{g}k_{\mu}k_{\nu} \, , \\
    f_{\mu\nu}= &C^{2}(\overline{f}^{*}_{\mu\nu}+\kappa_{f}\mathscr{h}_{\mu\nu}) \,, \quad \overline{f}^{}_{\mu\nu}=C^{2}\overline{f}^{*}_{\mu\nu} \,  , \quad \mathscr{h}_{\mu\nu} =  \phi_{f}l_{\mu}l_{\nu},
\end{align}
where $g_{\mu\nu}$ and $ f_{\mu\nu}$ are the full metrics and $\overline{g}_{\mu\nu}$ and $\overline{f}_{\mu\nu}$ are the background metrics. The conformal factor $C$ that appears in (\ref{type_b_ansatz_1}) is a dimensionless constant that is used in order to have flexibility when choosing the cosmological constants for the metrics. The perturbations to each background metrics are $h_{\mu\nu}$ and $\mathscr{h}_{\mu\nu}$ and $\phi_{g}$ and $\phi_{f}$ are their corresponding scalar functions. The vector $k_{\mu}$ is null with respect to the two metrics $g_{\mu\nu}$ and $\overline{g}_{\mu\nu}$ and the vector $l_{\mu}$ is null with respect to the two metrics $f_{\mu\nu}$ and $\overline{f}_{\mu\nu}$, i..e.
\begin{align}
\label{null_cond}
    g^{\mu\nu} k_{\mu} k_{\nu} = \,0 = \overline{g}^{\mu\nu} k_{\mu} k_{\nu}  \,, \quad 
    f^{\mu\nu} l_{\mu} l_{\nu} = & \, 0 = \overline{f}^{\mu\nu} l_{\mu} l_{\nu} \, . 
\end{align}
The covariant derivatives associated to the full metrics $g_{\mu\nu}$ and $f_{\mu\nu}$ by their respective compatibility conditions are $\,^{(g)}\nabla_{\mu}$ and $\,^{(f)}\nabla_{\mu}$ and for the background metrics $\overline{g}_{\mu\nu}$ and $\overline{f}_{\mu\nu}$, the associated operators are $\,^{(g)}\overline{\nabla}_{\mu}$ and $\,^{(f)}\overline{\nabla}_{\mu}$. This is equivalent to the expressions:
\begin{align}
\label{comp_cond_ads_b}
    \,^{(g)}\nabla_{\lambda}\, g_{\mu\nu}=0 \,, \quad \,^{(g)}\overline{\nabla}_{\lambda}\, \overline{g}_{\mu\nu}=0 \,, \qquad \,^{(f)}\nabla_{\lambda}\, f_{\mu\nu}=0 \,, \quad \,^{(f)}\overline{\nabla}_{\lambda}\, \overline{f}_{\mu\nu}=0\, .
\end{align}
For general relativity, it is a know result \cite{Stephani:2003tm} that if we have two vacuum spacetimes $(S_g,g_{\mu \nu})$ and $(\overline{S}_g, \overline{g}_{\mu \nu})$ related by a Kerr-Schild transformation, the null vector $k_{\mu}$ that appears in the transformation is geodesic with respect both metrics. We can extend this result for vacuum spacetimes in bigravity for the generalized Kerr-Schild ansatz (\ref{type_b_ansatz_1}) and for null vectors (\ref{null_cond}) by imposing that the tensors $V_{\mu\nu}$ and $\mathcal{V}_{\mu\nu}$ satisfy the following conditions:
\begin{align}
    \label{cond_geod_bg}
    V_{\mu\nu} \,k^{\mu}k^{\nu} = 0 \, , \qquad \mathcal{V}_{\mu\nu}\, l^{\mu}l^{\nu} = 0 \, ,
\end{align}
so that the null vectors $k_{\mu}$ and $l_{\mu}$ are geodesic with respect their full and background metrics, i.e., 
\begin{align}
\label{geod}
    k^{\mu} \,^{(g)} {\nabla}_{\mu} k_{\nu} = 0 =k^{\mu} \,^{(g)}\overline{\nabla}_{\mu} k_{\nu} \,, \qquad  l^{\mu} \,^{(f)}{\nabla}_{\mu} l_{\nu} = 0 = l^{\mu} \,^{(f)}\overline{\nabla}_{\mu} l_{\nu} \, .
\end{align}
Thus, for vacuum spacetimes $S_{g}$, $\overline{S}_{g}$ and $S_{f}$, $\overline{S}_{f}$ in bigravity related by a Kerr-Schild transformation, the condition in (\ref{cond_geod_bg}) implies that the null vectors are geodesic with respect to both their corresponding full and background metrics, as described in (\ref{geod}). We will provide further elaboration on this shortly. We raise and lower indices of the null vectors using their respective background metric:
\begin{align}
    \nonumber k^{\mu}=\overline{g}^{\mu\alpha}k_{\alpha} \, , \quad l^{\mu}={\overline{f}}^{\,\mu\alpha}l_{\alpha} \, .
\end{align}
For the background covariant derivatives, we raise and lower the indices using their respective metrics. Specifically, this reads as $\,^{(g)}\overline{\nabla}^{\mu}=\overline{g}^{\mu\alpha}\,^{(g)}\overline{\nabla}_{\alpha}, \, \,^{(f)}\overline{\nabla}^{\mu}=\overline{f}^{\,\mu\alpha}\,^{(f)}\overline{\nabla}_{\alpha}$. Similarly, in the case of the covariant derivatives of the full metrics, we have $\,^{(g)}\nabla^{\mu}=g^{\mu\alpha}\,^{(g)}\nabla_{\alpha}, \, \,^{(f)}\nabla^{\mu}=f^{\,\mu\alpha}\,^{(f)}\nabla_{\alpha}$. 

An important property of the Kerr-Schild ansatz is that the inverse metrics can be written as:
\begin{align}
    \nonumber g^{\mu \nu} =  \overline{g}^{\mu \nu} - \kappa_{g} \phi_{g} k^{\mu} k^{\nu} \,, \quad f^{\mu \nu} =  C^{-2}\left(\overline{f}^{*\,\mu \nu} - \kappa_{f} \phi_{f} l^{\mu} l^{\nu}\right) \,, 
\end{align}
which follows from the fact that both vectors $k_{\mu}$ and $l_{\mu}$ are null. From the compatibility condition for the full metrics we can derive the form of the Christoffel symbols:
\begin{align}
\label{Christoffels_bigravity} 
    \nonumber \,^{(g)}\nabla_{\lambda}g_{\mu\nu}=0 \quad \Longleftrightarrow \quad\,^{(g)}\Gamma^{\lambda}{}_{\mu\nu}&=\frac{1}{2}g^{\sigma\lambda}\left(\,^{(g)}\partial_{\mu}g_{\nu\sigma}+\,^{(g)}\partial_{\nu}g_{\mu\sigma}-\,^{(g)}\partial_{\sigma}g_{\mu\nu}\right)\,  ,\\
    \,^{(f)}\nabla_{\lambda}f_{\mu\nu}=0 \quad \Longleftrightarrow \quad \,^{(f)}\Gamma^{\lambda}{}_{\mu\nu}&=\frac{1}{2}f^{\sigma\lambda}\left(\,^{(f)}\partial_{\mu}f_{\nu\sigma}+\,^{(f)}\partial_{\nu}f_{\mu\sigma}-\,^{(f)}\partial_{\sigma}f_{\mu\nu}\right) \, . 
\end{align}
We can use the definition of the covariant derivative of the background metrics and apply it to the full metrics, which leads us to expressions
\begin{align}
    \nonumber \,^{(g)}\overline{\nabla}_{\lambda}\, g_{\mu\nu}=& \,^{(g)}\partial_{\lambda}\, g_{\mu\nu} - \,^{(g)}\overline{\Gamma}^{\rho}{}_{\lambda\mu}g_{\rho\nu}-\,^{(g)}\overline{\Gamma}^{\rho}{}_{\lambda\nu}g_{\mu\rho}=\,^{(g)}\overline{\nabla}_{\lambda}\,(\overline{g}_{\mu\nu}+\kappa_{g}h_{\mu\nu}) \\
    \nonumber =&\, \kappa_{g}\,^{(g)}\overline{\nabla}_{\lambda}\,h_{\mu\nu} \,,  \\
    \nonumber \,^{(f)}\overline{\nabla}_{\lambda}\, f_{\mu\nu}=&  \,\,^{(f)}\partial_{\lambda}\, f_{\mu\nu} - \,^{(f)}\overline{\Gamma}^{\rho}{}_{\lambda\mu}f_{\rho\nu}-\,^{(f)}\overline{\Gamma}^{\rho}{}_{\lambda\nu}f_{\mu\rho}=\,^{(f)}\overline{\nabla}_{\lambda}\,(\overline{f}_{\mu\nu}+C^{2}\kappa_{f}\mathscr{h}_{\mu\nu}) \\
    \nonumber =& \,  C^{2}\kappa_{f}\,^{(f)}\overline{\nabla}_{\lambda}\,\mathscr{h}_{\mu\nu} \,,
\end{align}
where we have also used (\ref{comp_cond_ads_b}) and the Christoffel symbols $\overline{\Gamma}^{\lambda}{}_{\mu\nu}$ for each of the background metrics . We can rewrite these results to obtain
\begin{align}
    \nonumber \,^{(g)}\partial_{\lambda}\, g_{\mu\nu} = \kappa_{g}\,^{(g)}\overline{\nabla}_{\lambda}\,h_{\mu\nu} + \,^{(g)}\overline{\Gamma}^{\rho}{}_{\lambda\mu}g_{\rho\nu} + \,^{(g)}\overline{\Gamma}^{\rho}{}_{\lambda\nu}g_{\mu\rho} \,,  \\
    \nonumber \,^{(f)}\partial_{\lambda}\, f_{\mu\nu} = C^{2}\kappa_{f}\,^{(f)}\overline{\nabla}_{\lambda}\,\mathscr{h}_{\mu\nu} + \,^{(f)}\overline{\Gamma}^{\rho}{}_{\lambda\mu}f_{\rho\nu} + \,^{(f)}\overline{\Gamma}^{\rho}{}_{\lambda\nu}f_{\mu\rho} \,. 
\end{align}
By substituting these last results in the Christoffel symbols of the full metrics (\ref{Christoffels_bigravity}), we can express them in terms of quantities that depend on the background metrics and on the perturbations. The result we obtain is given by
{\footnotesize
\begin{align}
    \label{christoffels}
    \nonumber \,^{(g)}\Gamma^{\lambda}{}_{\mu\nu} = &\,^{(g)}\overline{\Gamma}^{\lambda}{}_{\mu\nu}+\frac{\kappa_{g}}{2}\left[ \overline{g}^{\sigma\lambda}\left(\,^{(g)}\overline{\nabla}_{\mu}\,h_{\sigma\nu}+\,^{(g)}\overline{\nabla}_{\nu}\,h_{\mu\sigma}-\,^{(g)}\overline{\nabla}_{\sigma}\,h_{\mu\nu}\right)+\kappa_{g}h^{\sigma\lambda}\,^{(g)}\overline{\nabla}_{\sigma}\,h_{\mu\nu}\right] \, , \\
    \,^{(f)}\Gamma^{\lambda}{}_{\mu\nu} = & \,^{(f)}\overline{\Gamma}^{\lambda}{}_{\mu\nu}+\frac{C^{2}\kappa_{f}}{2}\left[ \overline{f}^{\sigma\lambda}\left(\,^{(f)}\overline{\nabla}_{\mu}\,\mathscr{h}_{\sigma\nu}+\,^{(f)}\overline{\nabla}_{\nu}\,\mathscr{h}_{\mu\sigma}-\,^{(f)}\overline{\nabla}_{\sigma}\,\mathscr{h}_{\mu\nu}\right)+C^{2}\kappa_{f}\mathscr{h}^{\sigma\lambda}\,^{(f)}\overline{\nabla}_{\sigma}\,\mathscr{h}_{\mu\nu}\right] \, .
\end{align}
}With these results we have an expression for the Christoffel symbols of the two metrics in bigravity in terms of the Christoffel symbols of the background fields and the perturbations to the background metrics. Some useful contractions of (\ref{christoffels}) can be done, for example, with two null vectors
\begin{align}
\label{christoffels_cont3}
    \nonumber & \,^{(g)}\Gamma^{\lambda}{}_{\mu\nu}k^{\mu}k^{\nu} =\,^{(g)}\overline{\Gamma}^{\lambda}{}_{\mu\nu}k^{\mu}k^{\nu},  && \,^{(g)}\Gamma^{\lambda}{}_{\mu\nu}k_{\lambda}k^{\nu}=\,^{(g)}\overline{\Gamma}^{\lambda}{}_{\mu\nu}k_{\lambda}k^{\nu}\, , \\
    & \,^{(f)}\Gamma^{\lambda}{}_{\mu\nu}l^{\mu}l^{\nu} =\,^{(f)}\overline{\Gamma}^{\lambda}{}_{\mu\nu}l^{\mu}l^{\nu}, && \,^{(f)}\Gamma^{\lambda}{}_{\mu\nu}l_{\lambda}l^{\nu}=\,^{(f)}\overline{\Gamma}^{\lambda}{}_{\mu\nu}l_{\lambda}l^{\nu}\, ,
\end{align}
which imply the following expressions
\begin{align}
\label{geodesic_vect}
    \nonumber & k^{\mu} \,^{(g)}{\nabla}_{\mu} k_{\nu} = k^{\mu} \,^{(g)}\overline{\nabla}_{\mu} k_{\nu}  \,, && k^{\mu} \,^{(g)}{\nabla}_{\mu} k^{\nu} = k^{\mu} \,^{(g)}\overline{\nabla}_{\mu} k^{\nu}  \,,  \\
    & l^{\mu} \,^{(f)}{\nabla}_{\mu} l_{\nu} = l^{\mu} \,^{(f)}\overline{\nabla}_{\mu} l_{\nu} \, ,
    && l^{\mu} \,^{(f)}{\nabla}_{\mu} l^{\nu} = l^{\mu} \,^{(f)}\overline{\nabla}_{\mu} l^{\nu}  \, .
\end{align}
On the other hand, the Ricci tensors for each of the metrics can be written as: 
\begin{align}
    \nonumber R_{\mu\nu}=\overline{R}_{\mu\nu}+\,^{(g)}\overline{\nabla}_{\alpha} D^{\alpha}{}_{\mu\nu}-D^{\alpha}{}_{\nu\beta}D^{\beta}{}_{\mu\alpha}\, , \qquad D^{\alpha}{}_{\mu\nu} \equiv \,^{(g)}\Gamma^{\alpha}{}_{\mu\nu}-\,^{(g)}\overline{\Gamma}^{\alpha}{}_{\mu\nu} \,,  \\
    \mathcal{R}_{\mu\nu}= \overline{\mathcal{R}}_{\mu\nu}+\,^{(f)}\overline{\nabla}_{\alpha}\mathcal{D}^{\alpha}{}_{\mu\nu}-\mathcal{D}^{\alpha}{}_{\nu\beta}\mathcal{D}^{\beta}{}_{\mu\alpha} \,,  \qquad  \mathcal{D}^{\alpha}{}_{\mu\nu} \equiv \,^{(f)}\Gamma^{\alpha}{}_{\mu\nu}-\,^{(f)}\overline{\Gamma}^{\alpha}{}_{\mu\nu} \, , 
\end{align}
and after some manipulations, the contraction of the Ricci tensors with null vectors results in:
\begin{align}
    \label{Ricci_geod}
    \nonumber &R_{\mu\nu}k^{\mu}k^{\nu}=-\kappa_{g}\,\phi_{g} \,g^{\mu\nu}\left(k^{\alpha}\,^{(g)}\nabla_{\alpha} k_{\mu}\right)\left(k^{\beta}\,^{(g)}\nabla_{\beta} k_{\nu}\right) \, , \\
    &\mathcal{R}_{\mu\nu}l^{\mu}l^{\nu}=-C^{2}\,\kappa_{f}\, \phi_{f} f^{\mu\nu}\left(l^{\alpha}\,^{(f)}\nabla_{\alpha} l_{\mu}\right)\left(l^{\beta}\,^{(f)}\nabla_{\beta} l_{\nu}\right)\, .
\end{align}
For general relativity, in vacuum, we would have that the null vector is geodesic with respect to the full and the background metrics. For bigravity, in vacuum, we have
\begin{align}
    \nonumber R_{\mu\nu}\,k^{\mu}k^{\nu}=Q_{\mu\nu}\,k^{\mu}k^{\nu} \, , \qquad \mathcal{R}_{\mu\nu}\,l^{\mu}l^{\nu}=\mathcal{Q}_{\mu\nu}\,l^{\mu}l^{\nu}\, ,
\end{align}
so that the null vectors are geodesic with respect to both their corresponding full and background metric if and only if the condition (\ref{cond_geod_bg}) is satisfied. In such case, (\ref{Ricci_geod}) is also zero and one can check that by considering (\ref{geodesic_vect}), (\ref{geod}) holds. A similar condition to (\ref{cond_geod_bg}) should also apply to the interaction tensor for the background spacetime.

Using the fact that the null vectors are geodesic (\ref{geod}), the Ricci tensors with mixed components for the generalized Kerr-Schild ansatz (\ref{type_b_ansatz_1}) in bigravity can be rewritten as:
\begin{align}
\label{Riccis_bigrav_ads_b}
    \nonumber R^{\mu}{}_{\nu}&=\overline{R}^{\mu}{}_{\nu}-\kappa_{g}\, h^{\mu\lambda} \overline{R}_{\lambda \nu}+\frac{\kappa_{g}}{2}\,^{(g)}\overline{\nabla}_{\lambda}\left[ \,^{(g)}\overline{\nabla}^{\mu}h^{\lambda}{}_{\nu}+\,^{(g)}\overline{\nabla}_{\nu}h^{\mu\lambda}-\,^{(g)}\overline{\nabla}^{\lambda}h^{\mu}{}_{\nu}\right] \, , \\
    \mathcal{R}^{\mu}{}_{\nu}&=\overline{\mathcal{R}}^{\mu}{}_{\nu}-C^{2}\kappa_{f} \, \mathscr{h}^{\mu\lambda} \overline{\mathcal{R}}_{\lambda\nu}+\frac{C^{2}\kappa_{f}}{2}\,^{(f)}\overline{\nabla}_{\lambda}\left[ \,^{(f)}\overline{\nabla}^{\mu}\mathscr{h}^{\lambda}{}_{\nu}+\,^{(f)}\overline{\nabla}_{\nu}\mathscr{h}^{\mu\lambda}-\,^{(f)}\overline{\nabla}^{\lambda}\mathscr{h}^{\mu}{}_{\nu}\right] \, .
\end{align}
These expressions are the analogue to the one presented in \cite{Stephani:2003tm} for general relativity. The Ricci tensors with the mixed indices are linear in the perturbations and the scalar functions. 

We are interested in studying the equations of bigravity in the context of the double copy, thus,  we will adopt the methodology presented in \cite{Carrillo-Gonzalez:2017iyj} and work with trace-reversed field equations. We could consider matter coupled to the fields, which from the equations of bigravity (\ref{bigrav_eom}) will lead us to
\begin{align}
    \nonumber R = -\left(Q+ \frac{{\kappa_{g}}^{2}}{2}T_{M}\right) \, , \quad \ \ \ \ \ \mathcal{R}= -\left(\mathcal{Q}+ \frac{{\kappa_{f}}^{2}}{2}\mathcal{T}_{M}\right),  
\end{align}
where ${T_{M}}^{\mu}{}_{\nu}$ and ${\mathcal{T}_{M}}^{\mu}{}_{\nu}$ are the (effective) matter energy-momentum tensors (depending on how one couples matter in bigravity). In these equations we have used the traces:
\begin{align}
    \nonumber Q\equiv Q^{\mu}{}_{\mu} \, , \quad \ \ \  T_{M} \equiv {T_{M}}^{\mu}{}_{\mu} \, , \quad \ \ \ \mathcal{Q}\equiv \mathcal{Q}^{\mu}{}_{\mu} \, , \quad  \ \ \ \mathcal{T}_{M} \equiv {\mathcal{T}_{M}}^{\mu}{}_{\mu} \, ,
\end{align}
which will permit us to write down the trace-reversed equations of bigravity as follows 
\begin{align}
\label{trace_rev_matter}
    \nonumber & R^{\mu}{}_{\nu} - Q^{\mu}{}_{\nu} + \frac{1}{2}Q \, \delta^{\mu}{}_{\nu} = \frac{{\kappa_{g}}^{2}}{2} {\check{T}_{M}{}}^{\mu}{}_{\nu} \,, \quad {\check{T}_{M}{}}^{\mu}{}_{\nu} \equiv {T_{M}}^{\mu}{}_{\nu}- \frac{1}{2}T_{M}  \, \delta^{\mu}{}_{\nu}  \, ,  \\
    & \mathcal{R}^{\mu}{}_{\nu} - \mathcal{Q}^{\mu}{}_{\nu} + \frac{1}{2}\mathcal{Q}\, \delta^{\mu}{}_{\nu} = \frac{{\kappa_{f}}^{2}}{2} {\check{\mathcal{T}}_M}^{\mu}{}_{\nu}\, , \quad {\check{\mathcal{T}}_M}^{\mu}{}_{\nu} \equiv {\mathcal{T}_{M}}^{\mu}{}_{\nu}- \frac{1}{2}\mathcal{T}_{M} \, \delta^{\mu}{}_{\nu}  \, , 
\end{align}
where we have defined the trace-reversed energy-momentum tensors ${\check{T}_{M}{}}^{\mu}{}_{\nu}$ and $ {\check{\mathcal{T}}_M}^{\mu}{}_{\nu}$. We can also consider localized sources in the equations, which we will be commenting on later. We will mostly be focusing on the vacuum case
\begin{align}
    R^{\mu}{}_{\nu} = Q^{\mu}{}_{\nu} - \frac{1}{2}Q\, \delta^{\mu}{}_{\nu}   \,, \quad  \ \ \ \ \ \mathcal{R}^{\mu}{}_{\nu} = \mathcal{Q}^{\mu}{}_{\nu} - \frac{1}{2}\mathcal{Q} \, \delta^{\mu}{}_{\nu} \, .
\end{align}
Finally, we can rewrite the previous equations using the expressions in (\ref{Riccis_bigrav_ads_b}) to obtain 
\begin{align}
\label{type_b_eom_gral}
    \nonumber \kappa_{g}\,^{(g)}\overline{\mathcal{E}}(h^{\mu}{}_{\nu}) - \kappa_{g}\,h^{\mu\lambda} \overline{R}_{\lambda \nu}+ \overline{R}^{\mu}{}_{\nu} \, = \,Q^{\mu}{}_{\nu} - \frac{1}{2}Q\, \delta^{\mu}{}_{\nu} \, , \\
    C^{2}\kappa_{f}\,^{(f)}\overline{\mathcal{E}}(\mathscr{h}^{\mu}{}_{\nu}) -C^{2} \kappa_{f} \, \mathscr{h}^{\mu\lambda} \overline{\mathcal{R}}_{\lambda\nu}+ \overline{\mathcal{R}}^{\mu}{}_{\nu} = \mathcal{Q}^{\mu}{}_{\nu} - \frac{1}{2}\mathcal{Q} \, \delta^{\mu}{}_{\nu} \, ,
\end{align}
where $^{(g)}\overline{\mathcal{E}}(h^{\mu}{}_{\nu})$ and $^{(f)}\overline{\mathcal{E}}(\mathscr{h}^{\mu}{}_{\nu})$ are the Lichnerowicz operators, which are defined by
\begin{align}
    \nonumber \,^{(g)}\overline{\mathcal{E}}(h^{\mu}{}_{\nu}) \equiv  \frac{1}{2}\,^{(g)}\overline{\nabla}_{\lambda}\left[ \,^{(g)}\overline{\nabla}^{\mu}h^{\lambda}{}_{\nu}+\,^{(g)}\overline{\nabla}_{\nu}h^{\mu\lambda}-\,^{(g)}\overline{\nabla}^{\lambda}h^{\mu}{}_{\nu}\right] \, , \\
    \nonumber \,^{(f)}\overline{\mathcal{E}}(\mathscr{h}^{\mu}{}_{\nu}) \equiv \frac{1}{2}\,^{(f)}\overline{\nabla}_{\lambda}\left[ \,^{(f)}\overline{\nabla}^{\mu}\mathscr{h}^{\lambda}{}_{\nu}+\,^{(f)}\overline{\nabla}_{\nu}\mathscr{h}^{\mu\lambda}-\,^{(f)}\overline{\nabla}^{\lambda}\mathscr{h}^{\mu}{}_{\nu}\right] \, . 
\end{align}
In principle, we could consider other assumptions or modifications in order to study the Kerr-Schild classical double copy of bigravity. For example, we could introduce a double Kerr-Schild ansatz or incorporate higher-curvature terms into the action for bigravity. An instance of the former is presented in \cite{Luna:2015paa}, where the authors considered the Taub-NUT solution in the context of the double copy. As for the latter, in \cite{Gialamas:2023lxj}, the authors explored an extension for bigravity involving quadratic curvature terms for each metric. Exploring the double copy in bigravity within these type of settings would be interesting. However, for now, we will focus on the generalized single Kerr-Schild ansatz as in (\ref{type_b_ansatz_1}) and the action for bigravity as in (\ref{bigrav_action}).

By specifying a generalized Kerr-Schild ansatz in (\ref{type_b_ansatz_1}), we can write down the trace-reversed vacuum bigravity equations of motion as in (\ref{type_b_eom_gral}). The aim is to calculate the right-hand side of these equations of motion, which depend on the interaction tensors and their traces. This will be done by determining a closed form for the powers of the interaction matrix, $\gamma^{\mu}{}_{\nu}$.  

\subsection{Linear interactions in proportional Kerr-Schild spacetimes}
In order to calculate the interaction part of equations of motion of bigravity, we aim to obtain the interaction tensors $Q^{\mu}{}_{\nu}$ and $\mathcal{Q}^{\mu}{}_{\nu}$ explicitly, so we need an expression for the interaction matrix $\gamma^{\mu}{}_{\nu}$ defined in (\ref{gamma_def}). For the general case in the generalized Kerr-Schild ansatz where $k_{\mu}\ne l_{\mu}$, the interaction side involves terms at most fourth order in the interaction matrix $(\gamma^{4})^{\mu}{}_{\nu}$, so we will have terms such as $h^{\mu\alpha} h_{\alpha}{}_{\nu}$, $h^{\mu}{}_{\alpha}\mathscr{h}^{\alpha}{}_{\nu}$, $\mathscr{h}^{\mu}{}_{\alpha}h^{\alpha}{}_{\nu}$, $\mathscr{h}^{\mu}{}_{\alpha}\mathscr{h}^{\alpha}{}_{\nu}$ and greater orders, as well as contractions of the background metrics with themselves $\overline{g}^{\mu\alpha}\overline{f}_{\alpha\nu}$ and with the null vector of the other metric, $\overline{f}_{\alpha\nu}h^{\mu\alpha}$ and $\overline{g}^{\mu\alpha} \mathscr{h}_{\alpha\nu}$. This can be seen when using (\ref{type_b_ansatz_1}) to compute, for example 
\begin{align}
\label{gamma_example_1}
     (\gamma^{2})^{\mu}{}_{\nu} \equiv \gamma^{\mu}{}_{\alpha} \gamma^{\alpha}{}_{\nu} = g^{\mu\alpha}f_{\alpha\nu} = C^{2}\left[\overline{g}^{\mu\alpha}\overline{f}^{*}_{\alpha\nu} - (\kappa_{g}\overline{f}^{*}_{\alpha\nu}h^{\mu\alpha}{}-\kappa_{f}\overline{g}^{\mu\alpha} \mathscr{h}_{\alpha\nu})-\kappa_{g}\kappa_{f}h^{\mu\alpha}\mathscr{h}_{\alpha\nu}\right]\, .
\end{align}
Thus we have an explicit expression for this power of the interaction matrix in terms of the perturbation and background metrics. We can also compute $(\gamma^{4})^{\mu}{}_{\nu}$ in this manner, but the difficulty comes when trying to find $\gamma^{\mu}{}_{\nu}$ in terms of the perturbations. Because we have an square root, we cannot calculate it directly with the metrics and our attempts to find $\gamma^{\mu}{}_{\nu}$ such that $\gamma^{\mu}{}_{\alpha} \gamma^{\alpha}{}_{\nu}$ equals (\ref{gamma_example_1}) have been unsuccessful. 

To solve this, the authors in \cite{Ayon-Beato:2015qtt} studied specific solutions in bigravity which do not present the difficulty of nonlinear terms. This family of solutions were presented originally in the following ansatz \cite{Ayon-Beato:2015qtt}
\begin{align}
\label{type_b_ansatz_2}
    \nonumber g_{\mu\nu}= &\overline{g}_{\mu\nu}+\kappa_{g}h_{\mu\nu} \, ,  \quad h_{\mu\nu} = \phi_{g}k_{\mu}k_{\nu} \, , \\
    f_{\mu\nu}\, = & \, C^{2}(\overline{g}_{\mu\nu}+\kappa_{f}\mathscr{h}_{\mu\nu}) \, , \quad \mathscr{h}_{\mu\nu} = \phi_{f}k_{\mu}k_{\nu} ,
\end{align}
where (\ref{type_b_ansatz_1}) is restricted to the case where $\overline{f}_{\mu\nu}=C^{2}\overline{g}_{\mu\nu}$ and $l_{\mu}= \, k_{\mu}$. The ansatz in (\ref{type_b_ansatz_2}) implies that $k^{\mu}l_{\mu}=0$, thus, it presents the advantage that the perturbations only contribute linearly to the powers $(\gamma^{n})^{\mu}{}_{\nu}$ and makes it easier to compute them, because terms like $h^{\mu\alpha}\mathscr{h}_{\nu\alpha}$ and higher-order terms vanish. If we want non-linear contributions, we need $k^{\mu}l_{\mu}\ne 0$. We will focus in the particular family of solutions in (\ref{type_b_ansatz_2}) and work with the case of linear interactions, leaving the discussion of non-linear interactions apart.

The ansatz in (\ref{type_b_ansatz_2}) ensures that the covariant derivatives with index down for the background metrics are the same $\overline{\nabla}_{\mu} \equiv \,^{(g)}\overline{\nabla}_{\mu}=\,^{(f)}\overline{\nabla}_{\mu}$, reflects the fact that background metrics are proportional and relates the null vectors and covariant derivatives with the index raised so that:
\begin{align}
    \nonumber & k^{\mu}=\, \overline{g}^{\mu\alpha}k_{\alpha} \, , \quad  l^{\mu}=\overline{f}^{\mu\alpha}k_{\alpha} \,  \quad &&\Rightarrow \quad l^{\mu}= C^{-2} k^{\mu} \, , \\
    \nonumber & \,^{(f)}\overline{\nabla}^{\mu}= \overline{f}^{\mu\alpha}\,^{(f)}\overline{\nabla}_{\alpha} = \,  C^{-2}\overline{g}^{\,\mu\alpha}\overline{\nabla}_{\alpha} \quad &&\Rightarrow \quad \,^{(f)}\overline{\nabla}^{\mu} = C^{-2} \overline{\nabla}^{\mu} \, ,
\end{align} 
where we have used $\overline{\nabla}^{\mu} \equiv \,^{(g)}\overline{\nabla}^{\mu} = \overline{g}^{\mu\alpha}\overline{\nabla}_{\alpha}\,$.

Moreover, the ansatz (\ref{type_b_ansatz_2}) permits a closed form for the powers of the interaction $(\gamma^{n})^{\mu}{}_{\nu}$, which was first noted also by the authors in \cite{Ayon-Beato:2015qtt}, and it was used to calculate the bigravity equations for this ansatz. We will give a quick overview of the procedure and restate the expressions for using them in this work. 

Given (\ref{type_b_ansatz_2}), the powers of the interaction $(\gamma^{n})^{\mu}{}_{\nu}$ take the form
\begin{align}
    \nonumber \left(\gamma^{n}\right)^{\mu}{ }_{v}=C^{n}\left[\delta^{\mu}{ }_{\nu}-\frac{n}{2}\left(\kappa_{g}\phi_{g}-\kappa_{f}\phi_{f}\right) k^{\mu} k_{\nu}\right] \, , \quad
    [\gamma^{n}]^{m} = (4 C^{n})^{m} \, ,
\end{align}
which can also be written as:
\begin{align}
\label{gamma_1st_ads_b_2}
    \left(\gamma^{n}\right)^{\mu}{ }_{v}=C^{n}\left[\delta^{\mu}{ }_{\nu}-\frac{n}{2}\left(\kappa_{g}h^{\mu}{}_{\nu}-C^{2}\kappa_{f}\mathscr{h}^{\mu}{}_{\nu}\right) \right] \, ,
\end{align}
where we have raised and lowered the indices according to
\begin{align}
    \nonumber h^{\mu}{}_{\nu}=\overline{g}^{\mu\alpha}\phi_{g}k_{\alpha}k_{\nu} \, , \quad \mathscr{h}^{\mu}{}_{\nu}=\overline{f}^{\mu\alpha}\phi_{f}k_{\alpha}k_{\nu} \, . 
\end{align}
With (\ref{gamma_1st_ads_b_2}), we can calculate the potential $\mathcal{U}[g,f]$ and $\tau^{\mu}{}_{\nu}$:
\begin{align}
    \nonumber \mathcal{U}[g,f] =-(P_{1}+C P_{2}) \, , \quad \tau^{\mu}{}_{\nu} =-C\left[P_{2}\, \delta^{\mu}{}_{\nu} + P_{0} \left(\kappa_{g}h^{\mu}{}_{\nu}-C^{2}\kappa_{f}\mathscr{h}^{\mu}{}_{\nu}\right)\right] \, , 
\end{align}
where we have utilized the following definitions:
\begin{align}
    \nonumber P_{0} &\equiv \frac{1}{2}\left(b_{1}+2Cb_{2}+C^{2}b_{3}\right) \, , \\
    \nonumber P_{1} &\equiv -\left(b_{0}+3 C b_{1}+3 C^{2} b_{2}+C^{3} b_{3}\right) \, , \\
    \nonumber P_{2} &\equiv -\left(b_{1}+3 C b_{2}+3 C^{2} b_{3}+C^{3} b_{4}\right)\, .
\end{align}
The other two things we are missing for (\ref{bigrav_eom}) are the determinants for each metric. We define $g\equiv \operatorname{det}\left(g_{\mu \nu}\right)$ and $f\equiv \operatorname{det}\left(f_{\mu \nu}\right)$ as the determinants for each of the full metrics and also $\overline{g}\equiv \operatorname{det}\left(\overline{g}_{\mu \nu}\right)$ and $\overline{f}\equiv \operatorname{det}\left(C^{2}\overline{g}_{\mu \nu}\right)$. By expanding the full metrics around the background metric we get:
\begin{align}
    \nonumber g=\overline{g}\, , \quad f=\overline{f}=C^{8}\overline{g} \,, \quad \Longrightarrow -\frac{\sqrt{-g}}{\sqrt{-f}}=-\frac{1}{C^4} \, ,
\end{align}
where we have used the fact that the vector $k_{\mu}$ is null.  Thus, the expressions for the right-hand side of the bigravity equations are given by
\begin{align}
    \nonumber \mathrm{V}^{\mu}{ }_{\nu} =& \tau^{\mu}{ }_{
    \nu}-\mathcal{U} \delta^{\mu}{ }_{\nu} \, \quad \Rightarrow \quad \mathrm{V}^{\mu}{ }_{\nu} = P_{1}\delta^{\mu}{}_{\nu} - C P_{0}\left(\kappa_{g}h^{\mu}{}_{\nu}-C^{2}\kappa_{f}\mathscr{h}^{\mu}{}_{\nu}\right), \\
    \nonumber \mathcal{V}^{\mu}{ }^{\nu} =& -\frac{\sqrt{-g}}{\sqrt{-f}} \tau^{\mu}{ }_{\nu} \, \quad \Rightarrow \quad \mathcal{V}^{\mu}{ }^{\nu} = \frac{P_{2}}{C^{3}} \delta^{\mu}{}_{\nu} +\frac{P_{0}}{C^{3}}\left(\kappa_{g}h^{\mu}{}_{\nu}-C^{2}\kappa_{f}\mathscr{h}^{\mu}{}_{\nu}\right) \, .
\end{align}
Further simplification of this result can be achieved by defining another set of constants:
\begin{align}
    \nonumber B_{0} = \frac{m^{2} \kappa_{g}}{\kappa^{2}}(P_{1}) \, , \quad \mathcal{B}_{0} = \frac{m^{2} \kappa_{f}}{\kappa^{2}}\left(\frac{P_{2}}{C^{5}}\right) \, , \quad B_{1} = \frac{m^{2} \kappa_{g}}{\kappa^{2}}(C P_{0}) \, , \quad  \mathcal{B}_{1} = \frac{m^{2} \kappa_{f}}{\kappa^{2}}\left(\frac{P_{0}}{C^{5}}\right) \, , 
\end{align}
which results in the following interaction tensors:
\begin{align}
\label{int_tensors_ads_2}
    \nonumber & Q^{\mu}{}_{\nu} = \kappa_{g}\left[B_{0} \delta^{\mu}{}_{\nu} - B_{1} \left(\kappa_{g}h^{\mu}{}_{\nu}-C^{2}\kappa_{f}\mathscr{h}^{\mu}{}_{\nu}\right)\right]  \, , \quad Q = 4\, \kappa_{g}\, B_{0} \, , \\
    & \mathcal{Q}^{\mu}{}_{\nu} = C^{2}\kappa_{f}\left[\,\mathcal{B}_{0}\delta^{\mu}{}_{\nu} + \mathcal{B}_{1} \left(\kappa_{g}h^{\mu}{}_{\nu}-C^{2}\kappa_{f}\mathscr{h}^{\mu}{}_{\nu}\right)\right]\, , \quad \mathcal{Q}=4\,C^{2}\kappa_{f}\,\mathcal{B}_{0}\, . 
\end{align}
We note that the quantities in (\ref{int_tensors_ads_2}) satisfy the conditions in (\ref{cond_geod_bg}). As an additional note, we observe that the constants $B_{1}$ and $\mathcal{B}_{1}$ are related by
\begin{align}
    \nonumber \frac{B_{1}}{C\kappa_{g}} = \frac{m^{2}}{\kappa^{2}} P_{0} = \frac{C^{5}}{\kappa_{f}}\mathcal{B_{1}}  \quad \Rightarrow \quad B_{1} = \frac{C^{6}\kappa_{g}}{\kappa_{f}} \mathcal{B}_{1}\, .
\end{align}
Consequently, the trace-reversed bigravity equations of motion, i.e, the double copy equations in terms of the spin-2 perturbation fields can be written as:
\begin{align}
\label{type_b_ads_eom_1}
    \nonumber \,^{(g)}\overline{\mathcal{E}}(h^{\mu}{}_{\nu}) - h^{\mu\lambda} \overline{R}_{\lambda \nu}+\frac{\overline{R}^{\mu}{}_{\nu}}{\kappa_{g}} = &\,  - B_{0}\,\delta^{\mu}{}_{\nu} - B_{1} \left(\kappa_{g}h^{\mu}{}_{\nu}-C^{2}\kappa_{f}\mathscr{h}^{\mu}{}_{\nu}\right) \, , \\
   \,^{(f)}\overline{\mathcal{E}}(\mathscr{h}^{\mu}{}_{\nu}) - \mathscr{h}^{\mu\lambda} \overline{\mathcal{R}}_{\lambda\nu}+\frac{\overline{\mathcal{R}}^{\mu}{}_{\nu}}{C^{2} \kappa_{f}} = & \, - \mathcal{B}_{0}\, \delta^{\mu}{}_{\nu} + \mathcal{B}_{1} \left(\kappa_{g}h^{\mu}{}_{\nu}-C^{2}\kappa_{f}\mathscr{h}^{\mu}{}_{\nu}\right)\, .
\end{align}
These are the equations presented in \cite{Ayon-Beato:2015qtt} which are consistent with the ansatz (\ref{type_b_ansatz_2}) and we will study them in the context of the double copy. These equations consist of a coupled system for the perturbations, which depend on quantities of the background metrics. With this result, we conclude this first part, where the objective was to rewrite the bigravity equations of motion from \cite{Ayon-Beato:2015qtt} in terms of the perturbations and was achieved in (\ref{type_b_ads_eom_1}). This was accomplished using  (\ref{type_b_ansatz_2}), which requires proportionality in the background metrics and the null vectors to be the same (or proportional). With these restrictions, we can construct the interaction terms and the equations of motion. In the following sections we will study the equations we obtained in the context of the double copy, where we will rewrite the equations and calculate the single and zeroth copy equations for the ansatz (\ref{type_b_ansatz_2}) in bigravity.

\subsection{Maximally symmetric spacetimes}
To get a prescription of the Kerr-Schild classical double copy in curved backgrounds for bigravity, we will study maximally symmetric spacetimes. This has been studied in \cite{Bahjat-Abbas:2017htu, Carrillo-Gonzalez:2017iyj} in the context of general relativity. A motivation to consider non-flat asymptotic spacetimes is that the interaction term in bigravity has a contribution from the cosmological constant for both metrics, so if we will study the double copy for bigravity, it is in our interest to study non-flat backgrounds.

In the context of general relativity, one would modify the Einstein equations to take into account the cosmological constant contribution. In bigravity, these terms arises naturally in the interaction between the metrics (the parameters $b_{0}$ and $b_{4}$ in the $\gamma$-formulation for example) in the tensors $Q^{\mu}{}_{\nu}$ and $\mathcal{Q}^{\mu}{}_{\nu}$ of the equations of motion in (\ref{bigrav_eom}). On the other hand, just as in general relativity we find that, for maximally symmetric spacetimes, the following hold for each background Ricci tensors:
\begin{align}
    \nonumber \overline{R}_{\mu\nu}=\Lambda_{g} \overline{g}_{\mu\nu} \, , \quad \overline{R}=4 \Lambda_{g}\, , \\
    \nonumber \overline{\mathcal{R}}_{\mu\nu}=\Lambda_{f} \overline{f}_{\mu\nu} \, , \quad \overline{\mathcal{R}}=4 \Lambda_{f}\, ,
\end{align}
where $\Lambda_{g}$ and $\Lambda_{f}$ are the cosmological constants for the background metrics $\overline{g}_{\mu\nu}$ and $\overline{f}_{\mu\nu}$. Considering the ansatz (\ref{type_b_ansatz_2}), for maximally symmetric spacetimes, we have $\overline{R}=C^{2}\overline{\mathcal{R}}$. We can perform some manipulations in the Ricci tensors in (\ref{Riccis_bigrav_ads_b}) to express them as
\begin{align}
\label{Riccis_bigrav_ads_dc_1}
    \nonumber & R^{\mu}{}_{\nu} = \overline{R}^{\mu}{}_{\nu}-\frac{\kappa_{g}}{2}\left(k_{\nu}\overline{\nabla}_{\lambda}F^{\lambda\mu}+\frac{1}{6}\overline{R}A^{\mu}k_{\nu}+X^{\mu}{}_{\nu}+Y^{\mu}{}_{\nu}\right), \\
    & \mathcal{R}^{\mu}{}_{\nu} = \overline{\mathcal{R}}^{\mu}{}_{\nu}-\frac{C^{2}\kappa_{f}}{2}\left(l_{\nu}\overline{\nabla}_{\lambda}\mathcal{F}^{\lambda\mu}+\frac{1}{6}\overline{\mathcal{R}}\mathcal{A}^{\mu}l_{\nu}+\mathcal{X}^{\mu}{}_{\nu}+\mathcal{Y}^{\mu}{}_{\nu}\right) \, ,
\end{align}
where we have used the following quantities:
\begin{align}
    \nonumber &X^{\mu}_{\nu} \equiv-\overline{\nabla}_{\nu}\left(k^{\mu}\overline{\nabla}_\lambda A^{\lambda}\right) \, , \quad Y^{\mu}{}_{\nu} \equiv F^{\rho\mu}\,\overline{\nabla}_{\rho}k_{\nu}-\overline{\nabla}_{\rho}\left(A^{\rho}\overline{\nabla}^{\mu}k_{\nu}-A^{\mu}\overline{\nabla}_{\rho}k_{\nu}\right) \, , \\
    \nonumber &\mathcal{X}^{\mu}_{\nu} \equiv-\overline{\nabla}_\nu\left(l^\mu\overline{\nabla}_\lambda \mathcal{A}^{\lambda}\right) \, , \quad \mathcal{Y}^{\mu}{}_{\nu} \equiv \mathcal{F}^{\rho\mu}\,\overline{\nabla}_{\rho}l_{\nu}-\overline{\nabla}_{\rho}\left(\mathcal{A}^{\rho}\,^{(f)}\overline{\nabla}^{\mu}l_{\nu}-\mathcal{A}^{\mu}\overline{\nabla}_{\rho}l_{\nu}\right) \, ,
\end{align}
and the abelian field and associated field strengths:
\begin{align}
\label{def_gauge_fields}
    \nonumber A_{\mu} \equiv \, & \phi_{g} k_{\mu} \, , \quad A^{\mu}\equiv \overline{g}^{\mu \alpha} A_{\alpha} \, , \quad  F^{\mu\nu}=\overline{\nabla}^{\mu}A^{\nu}-\overline{\nabla}^{\nu}A^{\mu} \, , \quad \\
    \mathcal{A}_{\mu}= \, & \phi_{f}k_{\mu} \, , \quad  \mathcal{A}^{\mu}\equiv \overline{f}^{\mu \alpha} \mathcal{A}_{\alpha} \, , \quad \mathcal{F}^{\mu\nu}=\,^{(f)}\overline{\nabla}^{\mu}\mathcal{A}^{\nu}-\,^{(f)}\overline{\nabla}^{\nu}\mathcal{A}^{\mu} \, .
\end{align}
In these results we are considering the general case where $l_{\mu}\ne k_{\mu}$. For the case where both null vectors are the same we substitute the Ricci tensors as written in (\ref{Riccis_bigrav_ads_dc_1}) into the trace-reversed equations. These results lead us to write the trace-reversed equations of bigravity for the ansatz in (\ref{type_b_ansatz_2}) as follows 
{\footnotesize\begin{align}
\label{dc_pert_b_1}
    \nonumber -\frac{1}{2}\left(k_{\nu}\overline{\nabla}_{\lambda}F^{\lambda\mu}+\frac{1}{6}\overline{R}A^{\mu}k_{\nu}- \frac{2}{\kappa_{g}} \overline{R}^{\mu}{}_{\nu}+X^{\mu}{}_{\nu}+Y^{\mu}{}_{\nu}\right) \,  =  - B_{0}\, \delta^{\mu}{}_{\nu} - B_{1} \left(\kappa_{g}h^{\mu}{}_{\nu}-C^{2}\kappa_{f}\mathscr{h}^{\mu}{}_{\nu}\right) \, , \\
    -\frac{1}{2}\left(k_{\nu}\overline{\nabla}_{\lambda}\mathcal{F}^{\lambda\mu}+\frac{1}{6}\overline{\mathcal{R}}\mathcal{A}^{\mu}k_{\nu}- \frac{2}{C^2\kappa_{f}} \overline{\mathcal{R}}^{\mu}{}_{\nu}+\mathcal{X}^{\mu}{}_{\nu}+\mathcal{Y}^{\mu}
    {}_{\nu}\right) \, = \, - \mathcal{B_{0}}\, \delta^{\mu}{}_{\nu} +\mathcal{B_{1}} \left(\kappa_{g}h^{\mu}{}_{\nu}-C^{2}\kappa_{f}\mathscr{h}^{\mu}{}_{\nu}\right)\, . 
\end{align}}
This is practically what is done in \cite{Carrillo-Gonzalez:2017iyj} with the trace-reversed Einstein equations. We are conveniently separating the contributions related to the Maxwell equations in curved spacetime and other contributions that depend on the perturbation and the background spacetime. We will follow the procedure in \cite{Carrillo-Gonzalez:2017iyj}  contract (\ref{dc_pert_b_1}) using Killing vectors to obtain the single and zeroth copy corresponding equations. We will use timelike Killing vectors for the stationary solutions and null Killing vectors for the time-dependent solutions we present in this article.

\subsection{Single and zeroth copy for generalized Kerr-Schild ansatz in bigravity}
We have presented the equations of bigravity (\ref{type_b_ads_eom_1}) for the ansatz in (\ref{type_b_ansatz_2}), which correspond to the double copy equations. We are interested in which equations we would obtain when applying the double copy procedure from \cite{Carrillo-Gonzalez:2017iyj} to the family of solutions in bigravity we are studying. Then, the question this paper seeks to address is to which single and zeroth copy equations the bigravity equations will be mapped to after applying the double copy approach. We will contract equations (\ref{dc_pert_b_1}) with Killing vectors $K^{\nu}$ for the metric $g_{\mu\nu}$ and $\mathcal{K}^{\nu}$ for the metric $f_{\mu\nu}$ in order to obtain the single and zeroth copy equations for the ansatz, as it was carried out in \cite{Carrillo-Gonzalez:2017iyj}. For stationary solutions we will use timelike Killing vectors and for time dependent solutions we will use null Killing vectors. This approach is used to obtain the correct sources in the single and zeroth copies. Remarkably, the method of employing Killing vectors to obtain the single and zeroth copy and the role that isometries play in the Kerr-Schild double copy was studied in \cite{Easson:2023dbk}.

Before making use of the trace-reversed equations rewritten as in (\ref{dc_pert_b_1}), we write some relations that will be useful for future convenience in the computations:
\begin{align}
    \nonumber \alpha \equiv \, K^{\rho}k_{\rho} = K_{\rho}k^{\rho} \, , \quad  \beta \equiv \, \mathcal{K^{\rho}}k_{\rho} = \mathcal{K}_{\alpha} l^{\alpha}  \, , \qquad
    C^{2} \beta =  \mathcal{K}_{\alpha} k^{\alpha} \, , \quad  C^{-2} \alpha = K_{\alpha}l^{\alpha}  \, ,
\end{align}
where these $\alpha$'s and $\beta$'s may be functions of the coordinates. We proceed by multiplying the equations in (\ref{dc_pert_b_1}) by $\frac{-2}{\alpha}K^{\nu}$ and $\frac{-2}{\beta}\mathcal{K}^{\nu}$ respectively to obtain the single copy equations:
{\footnotesize
\begin{align}
\label{eom_sc_1}
    \nonumber \overline{\nabla}_{\lambda}F^{\lambda\mu}+\frac{\overline{R}}{6}A^{\mu}+\frac{1}{\alpha}K^{\nu}\left(X^{\mu}{}_{\nu}+Y^{\mu}{}_{\nu}-\frac{2}{\kappa_{g}} \, \overline{R}^{\mu}{}_{\nu}\right) \, = \, &\frac{2 B_{1}}{\kappa_{g}\,\alpha}K^{\mu} + \frac{2 B_{2}}{\kappa_{g}} \left(\kappa_{g}A^{\mu}-C^{2}\kappa_{f}\mathcal{A}^{\mu}\right) \, , \\
    \overline{\nabla}_{\lambda} \mathcal{F}^{\lambda\mu}+\frac{\overline{\mathcal{R}}}{6}\mathcal{A}^{\mu}+\frac{1}{\beta}\mathcal{K}^{\nu}\left(\mathcal{X}^{\mu}{}_{\nu}+\mathcal{Y}^{\mu}{}_{\nu}-\frac{2}{C^{2}\kappa_{f}} \, \overline{\mathcal{R}}^{\mu}{}_{\nu}\right) \,
    = \, &\frac{2 \, \mathcal{B}_{1}}{C^{2} \kappa_{f}\, \beta}\mathcal{K}^{\mu}-\frac{2\, \mathcal{B}_{2}}{C^{2}\kappa_{f}} \left(\kappa_{g}A^{\mu}-C^{2}\kappa_{f}\mathcal{A}^{\mu}\right) \, .
\end{align}
}In order to calculate the zeroth copy equations, let's rewrite:
\begin{align}
\label{rwt_maxeqs_1}
    \overline{\nabla}_{\sigma} F^{\sigma \mu} =  k^{\mu}\overline{\nabla}^{2}\phi_{g}+\frac{1}{\alpha}Z^{\mu} \, , \quad \overline{\nabla}_{\sigma} \mathcal{F}^{\sigma \mu} =  l^{\mu} \,^{(f)}\overline{\nabla}^{2}\phi_{f} +\frac{1}{\beta}\mathcal{Z}^{\mu} \, ,
\end{align}
where we have used the laplacians for each metric $\overline{\nabla}^{2}\equiv \overline{\nabla}_{\sigma}\overline{\nabla}^{\sigma}$ and $\,^{(f)}\overline{\nabla}^{2}\equiv \overline{\nabla}_{\sigma}\,^{(f)}\overline{\nabla}^{\sigma}$ and also $Z^{\mu}$ and $\mathcal{Z^{\mu}}$, defined as
\begin{align}
    \nonumber Z^{\mu} \equiv \,& (K_{\lambda}k^{\lambda})\left[\overline{\nabla}_{\sigma}k^{\mu} \, \overline{\nabla}^{\sigma}\phi_{g}+\overline\nabla_{\sigma}(\phi_{g}\overline\nabla^{\sigma}k^{\mu}-\phi_{g}\overline\nabla^{\mu}k^{\sigma}-k^{\sigma}\overline\nabla^{\mu}\phi_{g})\right] \, , \\
    \nonumber \quad  \mathcal{Z}^{\mu} \equiv \, & (\mathcal{K}_{\lambda}l^{\lambda})\left[\overline{\nabla}_{\sigma}l^{\mu} \, \,^{(f)}\overline{\nabla}^{\sigma}\phi_{f}+\overline{\nabla}_{\sigma}\left(\phi_{f}\,^{(f)}\overline{\nabla}^{\sigma}l^{\mu}-\phi_{f}\,^{(f)}\overline\nabla^{\mu}l^{\sigma}-l^{\sigma}\,^{(f)}\overline\nabla^{\mu}\phi_{f}\right)\right] \, .
\end{align}
We can multiply both sides of (\ref{rwt_maxeqs_1}) by $\frac{1}{\alpha}K_{\mu}$ and $\frac{1}{\beta}\mathcal{K}_{\mu}$ to obtain
\begin{align}
    \nonumber \frac{1}{ \alpha}K_{\mu} \overline{\nabla}_{\sigma} F^{\sigma \mu} =  \overline{\nabla}^{2}\phi_{g}+\frac{1}{\alpha^2}K_{\mu}Z^{\mu} \, , \quad \frac{1}{\beta}\mathcal{K}_{\mu} \overline{\nabla}_{\sigma} \mathcal{F}^{\sigma \mu} =  \,^{(f)}\overline{\nabla}^{2}\phi_{f}+\frac{1}{\beta^2}\mathcal{K}_{\mu}\mathcal{Z}^{\mu}
\end{align}
and similarly manipulate the equation (\ref{eom_sc_1}) in order to arrive to:
{\scriptsize
\begin{align}
\label{eom_zc_1}
    \nonumber \overline{\nabla}^{2}\phi_{g}+\frac{\overline{R}}{6}\phi_{g}+\frac{1}{\alpha^2}K_{\mu}\left[K^{\nu}\left(X^{\mu}{}_{\nu}+Y^{\mu}{}_{\nu}-\frac{2}{\kappa_{g}} \, \overline{R}^{\mu}{}_{\nu}\right)+Z^{\mu}\right]  = &\, \frac{2 B_{0}}{\alpha^2}K_{\mu}K^{\mu} + 2 B_{1} \left(\kappa_{g}\phi_{g}-\kappa_{f}\phi_{f}\right) \, , \\
    \,^{(f)}\overline{\nabla}^{2}\phi_{f}+\frac{\overline{\mathcal{R}}}{6}\phi_{f}+\frac{1}{\beta^2}\mathcal{K}_{\mu}\left[\mathcal{K}^{\nu}\left(\mathcal{X}^{\mu}{}_{\nu}+\mathcal{Y}^{\mu}{}_{\nu}-\frac{2}{C^{2}\kappa_{f}} \overline{\mathcal{R}}^{\mu}{}_{\nu}\right)+\mathcal{Z}^{\mu}\right] 
    = &\, \frac{2\,  \mathcal{B}_{0}}{\beta^2}\mathcal{K}_{\mu}\mathcal{K}^{\mu} - 2\, C^{2}\,\mathcal{B}_{1} \left(\kappa_{g}\phi_{g}-\kappa_{f}\phi_{f}\right) \, ,
\end{align}}which are the zeroth copy equations of motion for the ansatz (\ref{type_b_ansatz_2}) in bigravity. What we have obtained is equivalent to the statement that the family of solutions we are studying in bigravity are mapped to (\ref{eom_sc_1}) by the single copy procedure and to (\ref{eom_zc_1}) by the zeroth copy procedure. This is true for the solutions we present, where the equations are simplified when considering stationary or time-dependent solutions.

At each level (double, single and zeroth copy) we would have a source in order for the double copy procedure to hold, i.e., we need sources in the gravity side to map to reasonable sources in the gauge side. As is stated in \cite{Carrillo-Gonzalez:2017iyj}, we can construct the correct sources of the single and zeroth copy sides from the gravity side source by using Killing vectors, so that if we start from the vacuum trace-reversed bigravity equations with energy-momentum tensors $T^{\mu}{}_{\nu}$ and $\mathcal{T^{\mu}{}_{\nu}}$ that source the solution
\begin{align}
\label{bigrav_eom_tr_emtensor}
    & R^{\mu}{}_{\nu} - Q^{\mu}{}_{\nu} + \frac{1}{2}Q \, \delta^{\mu}{}_{\nu} =\frac{{\kappa_{g}}^{2}}{2} {\check{T}}^{\mu}{}_{\nu} \,, \quad {\check{T}}^{\mu}{}_{\nu} \equiv T^{\mu}{}_{\nu}- \frac{1}{2} T \, \delta^{\mu}{}_{\nu}, \\
    & \mathcal{R}^{\mu}{}_{\nu} - \mathcal{Q}^{\mu}{}_{\nu} + \frac{1}{2}\mathcal{Q}\, \delta^{\mu}{}_{\nu} = \frac{{\kappa_{f}}^{2}}{2} \check{\mathcal{T}}^{\mu}{}_{\nu} \, , \quad \check{\mathcal{T}}^{\mu}{}_{\nu} \equiv \mathcal{T}^{\mu}{}_{\nu}- \frac{1}{2}\mathcal{T} \, \delta^{\mu}{}_{\nu} \, ,
\end{align}
which are not the same as ${T_{M}}^{\mu}{}_{\nu}$ and ${\mathcal{T}_{M}}^{\mu}{}_{\nu}$ in (\ref{trace_rev_matter}), and where we have used the trace-reversed energy-momentum tensors ${\check{T}}^{\mu}{}_{\nu}$ and $\check{\mathcal{T}}^{\mu}{}_{\nu}$, we can construct the sources for the single copy with the same procedure we applied before:
\begin{align}
    \nonumber J^{\mu}\equiv-\frac{- 2 K^{\nu}}{\alpha}\left(T^{\mu}{}_{\nu}-\frac{1}{2}T\delta^{\mu}{}_{\nu}\right) \, ,\quad \mathcal{J}^{\mu}\equiv-\frac{- 2 \mathcal{K}^{\nu}}{\beta}\left(\mathcal{T}^{\mu}{}_{\nu}-\frac{1}{2}\mathcal{T}\delta^{\mu}{}_{\nu}\right) \, ,
\end{align}
and also the zeroth copy sources as:
\begin{align}
    \nonumber j\equiv \frac{1}{\alpha}K_{\mu}J^{\mu} \, , \quad \ \ \ \ \ \mathscr{j}\equiv \frac{1}{\beta}\mathcal{K}_{\mu}\mathcal{J}^{\mu}\, .
\end{align}
To consider these terms we should add each source to the right-hand side of the trace-reversed bigravity equations and to the single and zeroth copy equations respectively. In the following examples, we will not focus on the sourcing of the solutions as we direct the study to the non-localized part of the equations.

With the above discussion, we have the general structure that we will utilize. In the next section we will focus on examining the equations obtained in the context of the double copy and we analyze them using particular solutions of bigravity. We will investigate whether the Kerr-Schild classical double copy holds, and if so, determine the equations satisfied by the corresponding fields.

\subsubsection{Reducing to Minkowski backgrounds}\label{ref_mink_back}
As an additional note before continuing with our analysis, we can reduce our results to Minkowski backgrounds by considering the case of vanishing cosmological constant for the two metrics and working in some coordinates such that $\eta_{\mu\nu}=\operatorname{diag}(-1,1,1,1)$. Given this case, the ansatz (\ref{type_b_ansatz_2}) takes the form:
\begin{align}
    \nonumber g_{\mu \nu} = \eta_{\mu \nu}+\kappa_{g}\phi_{g}k_{\mu}k_{\nu} \,, \quad \ \ \ \ f_{\mu \nu} = C^{2}(\eta_{\mu \nu}+\kappa_{f}\phi_{f}k_{\mu}k_{\nu}) \, 
\end{align}
and the equations in (\ref{type_b_ads_eom_1}) reduce to
{\small\begin{align}
    \nonumber R^{
    \mu}{}_{\nu}=\frac{1}{2}\partial_{\lambda}\left(\partial^{\mu}h^{\lambda}{}_{\nu}+\partial_{\nu}h^{\lambda\mu}-\partial^{\lambda}h^{\mu}{}_{\nu}\right) & =  - B_{0}\,\delta^{\mu}{}_{\nu} - B_{1} \left(\kappa_{g}h^{\mu}{}_{\nu}-C^{2}\kappa_{f}\mathscr{h}^{\mu}{}_{\nu}\right)  \, , \\
    \nonumber \mathcal{R}^{\mu}{}_{\nu}=\frac{1}{2}\partial_{\lambda}\left(\,^{(f)}\partial^{\mu}\mathscr{h}^{\lambda}{}_{\nu}+\,^{(f)}\partial_{\nu}\mathscr{h}^{\lambda\mu}-\,^{(f)}\partial^{\lambda}\mathscr{h}^{\mu}{}_{\nu}\right) & = \, - \mathcal{B}_{0}\,\delta^{\mu}{}_{\nu} + \mathcal{B}_{1} \left(\kappa_{g}h^{\mu}{}_{\nu}-C^{2}\kappa_{f}\mathscr{h}^{\mu}{}_{\nu}\right)\, ,
\end{align}}where $\partial^{\mu}=\eta^{\mu\nu}\partial_{\nu}=C^{2}\,^{(f)}\partial^{\lambda}$. Ricci tensors $R^{\mu}{}_{\nu}$ and $\mathcal{R}^{\mu}{}_{\nu}$ are linear in $h_{\mu\nu}$ and $\mathscr{h}_{\mu\nu}$ and this is an exact and coordinate-dependent result. For the single copy we have
\begin{align}
    \nonumber \nonumber \partial_{\lambda}F^{\lambda\mu}+\frac{1}{\alpha}K^{\nu}\left(X^{\mu}{}_{\nu}+Y^{\mu}{}_{\nu}\right) \, = & \frac{2\,B_{0}}{\alpha}K^{\mu} + 2\,B_{1} \left(\kappa_{g}A^{\mu}-C^{2}\kappa_{f}\mathcal{A}^{\mu}\right) \, , \\
    \nonumber \,^{(f)}\partial_{\lambda}\mathcal{F}^{\lambda\mu}+\frac{1}{\beta}\mathcal{K}^{\nu}\left(\mathcal{X}^{\mu}{}_{\nu}+\mathcal{Y}^{\mu}{}_{\nu}\right) \, = & \frac{2 \, \mathcal{B}_{0}}{\beta}\mathcal{K}^{\mu}-2\, \mathcal{B}_{1} \left(\kappa_{g}A^{\mu}-C^{2}\kappa_{f}\mathcal{A}^{\mu}\right) \, ,
\end{align}
and for the zeroth copy we get
\begin{align}
    \nonumber \partial^{2}\phi_{g}+\frac{1}{\alpha^2}K_{\mu}\left[K^{\nu}\left(X^{\mu}{}_{\nu}+Y^{\mu}{}_{\nu}\right)+Z^{\mu}\right] \, = \, & \frac{2 B_{0}}{\alpha^2}K_{\mu}K^{\mu} + 2 B_{1} \left(\kappa_{g}\phi_{g}-\kappa_{f}\phi_{f}\right) \, , \\
    \nonumber \,^{(f)}\partial^{2}\phi_{f}+\frac{1}{\beta^2}\mathcal{K}_{\mu}\left[\mathcal{K}^{\nu}\left(\mathcal{X}^{\mu}{}_{\nu}+\mathcal{Y}^{\mu}{}_{\nu}\right)+\mathcal{Z}^{\mu}\right] \, = \, & \frac{2\,  \mathcal{B}_{0}}{\beta^2}\mathcal{K}_{\mu}\mathcal{K}^{\mu} - 2\,C^{2}\,\mathcal{B}_{1} \left(\kappa_{g}\phi_{g}-\kappa_{f}\phi_{f}\right) \, .
\end{align}
We could simplify these results for the stationary case, where we can choose $k_{0}=\pm 1$ in order to identify the Kerr-Schild scalar field $\phi$ with the scalar potential in the gauge side \cite{Carrillo-Gonzalez:2017iyj}. Without loss of generality, we can proceed as in \cite{Monteiro:2014cda} and consider $k_{0}=-1$ as the null vector for both metrics. Then, if we take the Schwarzschild solution in cartesian coordinates, for example, we have
\begin{align}
    \nonumber X^{\mu}{}_{\nu} + \, Y^{\mu}{}_{\nu} = 0 \, , \quad \mathcal{X}^{\mu}{}_{\nu} + \mathcal{Y}^{\mu}{}_{\nu} = 0 \, , \quad Z^{\mu} \, = \, 0 \, , \quad \mathcal{Z}^{\mu} \, = \, 0 \, ,
\end{align}
with the constants $B_{0}, B_{1}, \mathcal{B}_{0}$ and $\mathcal{B}_{1}$ all equal to zero, which is consistent with the results of the following sections. With this choice of constants, we obtain a vacuum Maxwell equation in flat space and a massless Klein-Gordon equation, so that the formalism developed here provides analog results to the ones presented for the stationary case in \cite{Monteiro:2014cda} for Minkowski backgrounds in general relativity.

\section{Bigravity and the Kerr-Schild double copy} \label{BG_DC}
The double copy procedure has been applied to some exact solutions in general relativity. For our purposes, we are interested on studying bigravity solutions which can be written in a Kerr-Schild form. In this context of Kerr-Schild solutions, where the background is the Minkowski space, the double copy procedure has been applied to certain black holes like the Schwarzschild and the Kerr solution in \cite{Monteiro:2014cda}. For curved backgrounds some solutions have been studied in \cite{Carrillo-Gonzalez:2017iyj} like the Schwarzchild, Kerr and charged solutions in (A)dS spacetime, along with some wave solutions, also to the Taub-NUT solution in \cite{Luna:2015paa}. For more examples of the double copy in curved backgrounds, see \cite{Armstrong:2020woi,Albayrak:2020fyp,Liang:2023zxo,Lipstein:2023pih}. We will study some analogues of these solutions in general relativity in the framework of bigravity and the double copy.

We will focus on solutions such that the values of the coefficients $B_{0}$ and $\mathcal{B}_{0}$ are fixed by the diagonal contributions to the equations, which come from Ricci background terms $\overline{R}^{\mu}{}_{\nu}$ and $\overline{\mathcal{R}}^{\mu}{}_{\nu}$. This is our first restriction to start studying bigravity solutions in the context of the double copy and holds for the stationary and time-dependent solutions we study. In addition to this condition, we will have another condition which also holds for both type of solutions. We write both conditions as follows:
\begin{align}
\label{sols_dc_1}
    \nonumber & \overline{R}^{\mu}{}_{\nu}= \Lambda_{g}\delta^{\mu}{}_{\nu} \, , \quad  \overline{\mathcal{R}}^{\mu}{}_{\nu}=\Lambda_{f} \delta^{\mu}{}_{\nu} \,, \quad  B_{0}= -\frac{\Lambda_{g}}{\kappa_{g}}\, ,\quad  \mathcal{B}_{0}=-\frac{\Lambda_{f}}{C^{2}\,\kappa_{f}} \, , \\
    \nonumber \Rightarrow  \ \ \ \ \ \quad &  \frac{\overline{R}^{\mu}{}_{\nu}}{\kappa_{g}} + B_{0}\,\delta^{\mu}{}_{\nu}\, = 0 \, ,  \quad  \frac{ \overline{\mathcal{R}}^{\mu}{}_{\nu}}{C^{2}\,\kappa_{f}} + \mathcal{B}_{0}\, \delta^{\mu}{}_{\nu}\, = 0 \, , \\
    & \frac{1}{\alpha^2}K_{\mu}Z^{\mu} \, = \, \, - \frac{\overline{R}}{6}\phi_{g} \, , \quad \frac{1}{\beta^2}\mathcal{K}_{\mu}\mathcal{Z}^{\mu} \, = \, -\frac{\overline{\mathcal{R}}}{6}\phi_{f} \, .
\end{align}
Equations in (\ref{sols_dc_1}) hold for the stationary and time-dependent solutions in vacuum we study in this paper. When we restrict the values of $B_{0}, \mathcal{B}_{0},B_{1}$ and $\mathcal{B}_{1}$ we are fixing the values of the $b_{k}$ in $P_{0},P_{1}$ and $P_{2}$. In the presence of an external energy-momentum tensor, such as when one of the metrics is a Kerr-Newman-(A)dS solution, the last condition in (\ref{sols_dc_1}) is modified for the metric(s) coupled to matter. We will make a comment on this later, but for now we will focus on both types of vacuum solutions we are studying.

From the results of the previous section, we have derived the structure needed to study the Kerr-Schild double copy in bigravity. Expanding on this concept, the classical Kerr-Schild double copy in bigravity, as we refer to it, is an extension of the Kerr-Schild double copy in general relativity  \cite{Monteiro:2014cda} for two metric fields. Given that bigravity involves two metrics and thus, two spin-2 fields, we apply the double copy procedure (single copy  procedure, in this case) to each of the metrics separately to define two single copy fields, one for each metric. This is done by replacing one of the null vectors $k_{\mu}$ from the perturbations $h_{\mu\nu}$ and $\mathscr{h}_{\mu\nu}$ with color factors $c^{a}$ and $\mathscr{c}^{a}$ in the gauge theory to obtain the fields
\begin{align}
\label{sing_replace}
    A^{a}_{\mu}=c^{a}\, \phi_{g}\,k_{\mu}\, , \quad \frac{\kappa_{g}}{2} \rightarrow g_{g} \, , \qquad \mathcal{A}^{a}_{\mu}=\mathscr{c}^{a}\, \phi_{f}\,k_{\mu} \,, \quad \frac{\kappa_{f}}{2} \rightarrow g_{f} \, ,
\end{align}
which are just the non-Abelian version of the fields defined in (\ref{def_gauge_fields}). We also need to change the coupling constants of the gravitational theory $\kappa_{g}$, $\kappa_{f}$ to the gauge theory coupling constants $g_{g}$, $g_{f}$ and the gravity sources to color sources. If we repeat this process and replace another factor of $k_{\mu}$ for the respective color factor for each metric we arrive to the zeroth copy, which is obtained upon the substitutions:
\begin{align}
\label{zero_replace}
    {\Phi_{g}}^{aa'}=c^{a}c^{a'} \phi_{g}\, , \quad g_{g} \rightarrow y_{g} \, , \qquad {\Phi_{f}}^{aa'}=\mathscr{c}^{a}\mathscr{c}^{a'}\phi_{f}\,, \quad g_{f} \rightarrow y_{f} \, ,
\end{align}
where $y_{g}$ and $y_{f}$ are the coupling constants for the biadjoint theory. For our purposes, we will focus on the fields without the color factors and refer to them as the single copy $A_{\mu}\equiv \phi_{g}k_{\mu}$, $\mathcal{A}_{\mu}\equiv \phi_{f}k_{\mu}$ and the zeroth copy $ \phi_{g},$ $\phi_{f}$. Even though we have not justified the use of the Kerr-Schild double copy procedure in bigravity at the level of scattering amplitudes, we will work with the equations we have derived from the bigravity field equations and check how this classical Kerr-Schild double copy procedure works for the two metrics in bigravity, and if it relates to the equations obtained in general relativity.
\subsection{Stationary solutions}
The stationary solutions we presented here are basically black holes such as Schwarzschild, Kerr and their generalizations for maximally symmetric spacetimes. One distinction between the stationary and the time-dependent solutions that can be seen from the equations is that the following holds for these stationary solutions:
\begin{align}
\label{stat_constraints}
    \frac{\overline{R}}{6}h^{\mu}{}_{\nu} + X^{\mu}{}_{\nu}+Y^{\mu}{}_{\nu} =  0 \,, \quad \frac{\overline{\mathcal{R}}}{6}\mathscr{h}^{\mu}{}_{\nu}+\mathcal{X}^{\mu}{}_{\nu}+\mathcal{Y}^{\mu}{}_{\nu}= \, 0 \, , \quad  B_{1} = 0 \,, \quad \mathcal{B}_{1} = 0 \, . 
\end{align}
These conditions are modified if we couple matter to one of the metrics, as we will see in the Kerr-Newman-(A)dS solution we present later. Because the constants $B_{1}$ and $\mathcal{B}_{1}$ are zero for these solutions, the interaction terms $Q^{\mu}{}_{\nu}$ and $\mathcal{Q}^{\mu}{}_{\nu}$ will not have linear contributions in the perturbations. The interaction between the two metrics results in an effective cosmological constant term that depends only on the cosmological constant of the metrics.
Without the presence of an external field, in this case we obtain a couple of Maxwell and two conformally coupled equations:
\begin{align}
\label{stat_eqs}
    \nonumber  \,^{(g)}\overline{\mathcal{E}}(h^{\mu}{}_{\nu}) -\overline{R}^{\lambda}{}_{\nu}\, h^{\mu}{}_{\lambda} = 0 \, \quad \xrightarrow{SC} & \quad \overline{\nabla}_{\lambda}F^{\lambda\mu} = \, 0 \quad \xrightarrow{ZC} \quad \overline{\nabla}^{2}\phi_{g}-\frac{\overline{R}}{6}\phi_{g} = \, 0 \, , \\
    \,^{(f)}\overline{\mathcal{E}}(\mathscr{h}^{\mu}{}_{\nu}) - \overline{\mathcal{R}}^{\lambda}{}_{\nu}\mathscr{h}^{\mu}{}_{\lambda}  = 0 \, \quad \xrightarrow{SC} & \quad  \overline{\nabla}_{\lambda}\mathcal{F}^{\lambda\mu} = \,  0 \quad \xrightarrow{ZC} \quad  \,^{(f)}\overline{\nabla}^{2}\phi_{f}-\frac{\overline{\mathcal{R}}}{6}\phi_{f}  =  \,0 \, .
\end{align}
The equations we obtained for these stationary solutions correspond to massless spin-2, spin-1 and spin-0 fields. Then, based on this result, the black holes we are presenting do no possesses massive propagating degrees of freedom. At the level of the double copy, we have two decoupled equations for massless spin-2 fields, so we have $2+2=4$ degrees of freedom, corresponding to the two propagating degrees of freedom of each of the fields. For the single copy we have two decoupled massless spin-1 fields, so we have $2+2=4$ degrees of freedom. For the zeroth copy we have $1+1=2$ total degrees of freedom that correspond to each of the massless scalar fields. We proceed to present a couple of stationary-solution examples in bigravity in curved backgrounds.

\subsubsection{Example: Schwarzschild-(A)dS solution in bigravity}\label{SchwdAdS_example}

Spherically solutions for bimetric gravity have been found in \cite{Comelli:2011wq,Volkov:2012wp} (see also \cite{Babichev:2015xha})  and here we will work with the Schwarzschild-(A)dS solution in bigravity. In global static coordinates we can write the background (A)dS metric as
\begin{align}
    \nonumber d\overline{s}^{2}=-\left(1-\frac{\Lambda r^2}{3}\right)dt^2+\left(1-\frac{\Lambda r^2}{3}\right)^{-1}dr^2+r^2d\Omega^2
\end{align}
and the full metric is constructed with
\begin{align}
\label{schwd_ads_1}
    \phi_{g} =\frac{\kappa_{g}}{2}\frac{m_{1}}{4\,\pi\,r} \, , \quad \phi_{f}=\frac{\kappa_{f}}{2}\frac{m_{2}}{4\,\pi\,r} \, ,  \quad k_{\mu}=\left(1, \frac{1}{1-\frac{\Lambda r^2}{3}},0,0\right) \,  ,
\end{align}
so both metrics have the same cosmological constant $\Lambda_{g} = C^{2}\, \Lambda_{f} = \Lambda$ but different masses.  In this case the interaction constants take the values
\begin{align}
\label{ads_schwd_const}
    \nonumber P_{0}  =0 \,, \quad  & \Rightarrow \quad B_{1}=  \mathcal{B}_{1} = 0 \, , \\ P_{1}=\frac{{\kappa_{f}}^{2}P_{2}}{{C\,\kappa_{g}}^{2}}=-\frac{\Lambda \, \kappa^{2}}{m^{2}{\kappa_{g}}^{2}}\, \quad & \Rightarrow \quad \kappa_{g}\,B_{0}=C^{4}\,\kappa_{f}\, \mathcal{B}_{0}= -\Lambda.
\end{align}
Equation (\ref{ads_schwd_const}) implies $Q^{\mu}{}_{\nu}=- \Lambda \delta^{\mu}{}_{\nu} \, , \,  \mathcal{Q}^{\mu}{}_{\nu}= - \,C^{-2}\,\Lambda \delta^{\mu}{}_{\nu}$ so we can calculate explicitly using the metric and the interaction constants:
\begin{align}
    \nonumber R^{\mu}{}_{\nu}=\Lambda \delta^{\mu}{}_{\nu}=Q^{\mu}{}_{\nu}-\frac{1}{2}Q\, \delta^{\mu}{}_{\nu}\, , \quad \ \ \ \ \ \mathcal{R}^{\mu}{}_{\nu}=\frac{\Lambda}{C^{2}} \delta^{\mu}{}_{\nu} =\mathcal{Q}^{\mu}{}_{\nu}-\frac{1}{2}\mathcal{Q}\, \delta^{\mu}{}_{\nu} \, ,
\end{align}
where the equations in (\ref{stat_constraints}) hols for this solution. With this we have verified that the Schwarzschild-(A)dS is a solution to bigravity. The equations for the perturbations and the double copy fields are given by
\begin{align}
    \nonumber  \,^{(g)}\overline{\mathcal{E}}(h^{\mu}{}_{\nu}) -\overline{R}^{\lambda}{}_{\nu}\, h^{\mu}{}_{\lambda} & = 0 \, , \quad  h_{\mu\nu}=\frac{\kappa_{g}}{2}\frac{m_{1}}{4\,\pi\,r} k_{\mu}k_{\nu}, \\
   \,^{(f)}\overline{\mathcal{E}}(\mathscr{h}^{\mu}{}_{\nu}) - \overline{\mathcal{R}}^{\lambda}{}_{\nu}\mathscr{h}^{\mu}{}_{\lambda}  & = 0 \, , \quad \mathscr{h}_{\mu\nu}=\frac{\kappa_{f}}{2}\frac{m_{2}}{4\,\pi\,r} k_{\mu}k_{\nu}\, .
\end{align}
Now we see how the single and zeroth copy behave. We use the timelike Killing vector of the Schwarzschild metric and also the abelian fields and their field strengths as defined in (\ref{def_gauge_fields}). By also using the replacements in (\ref{sing_replace}) and (\ref{zero_replace}), the single copy equations for the Schwarzschild-(A)dS solution are:
\begin{align}
\label{sc_ads_schwd}
    \overline{\nabla}_{\lambda}F^{\lambda\mu} = 0 \, ,\quad A_{\mu}=g_{g}\frac{m_{1}}{4\,\pi\,r} k_{\mu} \, , \qquad \overline{\nabla}_{\lambda}\mathcal{F}^{\lambda\mu} = 0 \, , \quad \mathcal{A}_{\mu}= g_{f}\frac{m_{1}}{4\,\pi\,r} k_{\mu}\, .
\end{align}
We also need to map the gravitational sources $m_{1}$ and $m_{2}$ to color sources, but for simplicity we will stick to sources. By rewriting and contracting (\ref{sc_ads_schwd}) with the corresponding Killing vector, we note that for this solution we obtain that the zeroth copy equations are
$$
\overline{\nabla}^{2}\phi_{g}-\frac{\overline{R}}{6}\phi_{g} = \, 0\,  , \ \ \ \ \  \phi_{g}=y_{g}\frac{m_{1}}{4\,\pi\,r} \, ,
$$
\begin{equation}
\label{zc_ads_schwd}    
^{(f)}\overline{\nabla}^{2}\phi_{f}-\frac{\overline{\mathcal{R}}}{6}\phi_{f} = \,0 \, , \ \ \ \ \ \phi_{f}=y_{f}\frac{m_{2}}{4\,\pi\,r} \, . 
\end{equation}
We note that in equation (\ref{sc_ads_schwd}), we obtained a pair of Maxwell equations, while in equation  (\ref{zc_ads_schwd}) we obtained two of conformally coupled scalar field equations. At the single and the zeroth copy level, both equations are defined in the backgrounds $\overline{g}_{\mu\nu}$ and $\overline{f}_{\mu\nu}$ respectively. These equations coincide with what was given in (\ref{stat_eqs}). 

If we consider a Schwarzschild solution in bigravity in Eddington-Finkelstein coordinates we have that
\begin{align}
    \nonumber  d\overline{s}^{2}=-dt^{2}+dr^{2}+r^{2}d\Omega^2\, , \quad \phi_{g} =\frac{\kappa_{g}}{2}\frac{m_{1}}{4\,\pi\,r} \, , \quad \phi_{f}=\frac{\kappa_{f}}{2}\frac{m_{2}}{4\,\pi\,r} \, ,  \quad k_{\mu}=(-1,-1,0,0) \, ,
\end{align}
where the cosmological constants are zero and the corresponding interaction constants are all zero $P_{0}  = P_{1}=P_{2}=0$. The equations that hold are also the ones in (\ref{stat_eqs}) with vanishing background scalar curvature $R=\mathcal{R}=0$. This formalism also holds when we use coordinates such that $\overline{g}_{\mu\nu}=\eta_{\mu\nu}=\operatorname{diag}(-1,1,1,1)$, reducing to the results in (\ref{ref_mink_back}) for flat backgrounds.

\subsubsection{Example: Kerr-(A)dS solution in bigravity}\label{KAdS_example}
The Kerr solution in bigravity was initially introduced in \cite{Babichev:2014tfa} as a rotating solution with different masses but the same rotation parameter. In this study, we will make use of the generalization to the Kerr-(A)dS solution in bigravity that was originally derived in \cite{Ayon-Beato:2015qtt}, utilizing the Kerr-Schild ansatz. This approach enables the expression of the solution in Kerr-Schild form using ellipsoidal coordinates. The background line element is the (A)dS metric and takes the form
\begin{align}
    \nonumber d\overline{s}^{2}=-\frac{\Delta}{\Omega}\, \xi \,  dt^{2}+\frac{\Sigma \,  dr^{2}}{(r^{2}+a^{2})\, \xi}+\frac{\Sigma \, d\theta^{2}}{\Delta}+\frac{(r^{2}+a^{2})}{\Omega}\sin^{2}\theta d\varphi^{2}\, , 
\end{align}
where we have used
\begin{align}
    \nonumber \xi \equiv1-\frac{\Lambda r^2}{3},\quad\Sigma\equiv r^2+a^2\cos^2\theta,\quad\Delta\equiv1+\frac{\Lambda}{3}a^2\cos^2\theta,\quad\Omega\equiv1+\frac{\Lambda}{3}a^2.
\end{align}
For this solution, both metrics have the same cosmological constant $\Lambda_{g} = C^{2}\, \Lambda_{f} = \Lambda$ and rotation parameter $a$ but different masses. The null vector and scalar functions for this case are
\begin{align}
    \nonumber \phi_{g} =\frac{\kappa_{g}}{2}\frac{m_{1}r}{4\,\pi\,\Sigma} \, , \quad \phi_{f}=\frac{\kappa_{f}}{2}\frac{m_{2}r}{4\,\pi\,\Sigma}  \, ,  \quad k_{\mu}=\left(\frac{\Delta}{\Omega}, \frac{\Sigma}{(r^{2}+a^{2})\,\xi},0,-\frac{a \sin^{2}\theta}{\Omega}\right) \, ,
\end{align}
and  the constants $B_{0}, B_{1}, \mathcal{B}_{0}$ and $\mathcal{B}_{1}$ are the same as in (\ref{ads_schwd_const}). This implies $Q^{\mu}{}_{\nu}=- \Lambda\,\delta^{\mu}{}_{\nu} \, , \,  \mathcal{Q}^{\mu}{}_{\nu}= - \, C^{-2}\Lambda \,\delta^{\mu}{}_{\nu}$ so that bigravity equations hold, i.e.
\begin{align}
    \nonumber R^{\mu}{}_{\nu}=\Lambda \delta^{\mu}{}_{\nu}=Q^{\mu}{}_{\nu}-\frac{1}{2}Q\, \delta^{\mu}{}_{\nu}\, , \quad \mathcal{R}^{\mu}{}_{\nu}=\frac{\Lambda}{C^{2}} \delta^{\mu}{}_{\nu} =\mathcal{Q}^{\mu}{}_{\nu}-\frac{1}{2}\mathcal{Q}\, \delta^{\mu}{}_{\nu} \, ,
\end{align}
and (\ref{stat_constraints}) is  also fulfilled. We use the same Killing vector as in the Schwarzschild solution. The behaviour of the double, single and zeroth copy equations of this solution are the same as in the Schwarzschild-(A)dS solution, so that the equations we obtain for the Kerr-(A)dS solution are the ones in (\ref{stat_eqs}) for the fields:
\begin{align}
    \nonumber h_{\mu\nu}=\frac{\kappa_{g}}{2}\frac{m_{1}r}{4\,\pi\,\Sigma}k_{\mu}k_{\nu}\, , \quad A_{\mu}=g_{g}\frac{m_{1}r}{4\,\pi\,\Sigma}k_{\mu} \, , \quad \phi_{g} & =y_{g}\frac{m_{1}r}{4\,\pi\,\Sigma}\, ,  \\
     \mathscr{h}_{\mu\nu}=\frac{\kappa_{f}}{2}\frac{m_{2}r}{4\,\pi\,\Sigma}k_{\mu}k_{\nu}\, , \quad \mathcal{A}_{\mu}=g_{f}\frac{m_{2}r}{4\,\pi\,\Sigma}k_{\mu}\, , \quad \phi_{f} & =y_{f}\frac{m_{2}r}{4\,\pi\,\Sigma}\, .
\end{align}
Considering the case of background spacetimes with vanishing cosmological constant, the Kerr solution in bigravity \cite{Babichev:2014tfa} in ellipsoidal coordinates is given by
\begin{align}
    \nonumber & d\overline{s}^{2}=-dt^{2}+\left(r^{2}+a^{2}\right) \sin ^{2} \theta \,  d\phi^{2}+\frac{\Sigma}{r^{2}+a^{2}}d r^{2}+\Sigma d\theta^{2}\, , \\
    \nonumber & \phi_{g} =\frac{\kappa_{g}}{2}\frac{m_{1}r}{4\,\pi\,\Sigma} \, , \quad \phi_{f}=\frac{\kappa_{f}}{2}\frac{m_{2}r}{4\,\pi\,\Sigma} \, ,  \quad k_{\mu}=\left(1, \frac{\Sigma}{r^{2}+a^{2}},0,-a \sin^{2}\theta\right) \, ,
\end{align}
and the interaction constants $P_{0}= P_{1}=P_{2}=0$ are zero.

\subsubsection{Example: A KN-(A)dS interacting with a Kerr-(A)dS}

We can also explore the scenario when only one of the metrics in bigravity is coupled to a matter field. Coupling the Maxwell field to one of the Kerr black holes was originally discussed in \cite{Babichev:2014tfa} for the asymptotically flat case. Here, we will examine the asymptotically (A)dS case, specifically focusing on a Kerr-Newman-(A)dS solution within the framework of bigravity, which was first found in \cite{Ayon-Beato:2015qtt} by leveraging similarities with the matterless case. In this setup, the metric $g_{\mu\nu}$ is coupled to the electromagnetic matter field such that the scalar functions and the vector potential for this solution take the form:
\begin{align}
\label{knads_sol}
    \nonumber \phi_{g} = \frac{\kappa_{g}}{2}\frac{m_{1}r}{4\, \pi \, \Sigma}+\phi_{M} \, , \quad  \phi_{M}\equiv -\frac{\kappa_{g}}{2}\frac{q^{2}}{8\,\pi\,\Sigma} \, , \quad \phi_{f} = \frac{\kappa_{f}}{2}\frac{m_{2}r}{4\,\pi\,\Sigma}\, , \quad A_{M\,\mu}\equiv \frac{q\,r}{\Sigma}k_{\mu}\, ,
\end{align}
where the null vector $k_{\mu}$ is the same as the one used in the Kerr-(A)dS solution. $\phi_{M}$ is the scalar function that accounts for the electric charge $q$ of the solution and $A_{M\mu}$ is the field that corresponds the matter content of the theory. The energy momentum tensor $T_{M\,\mu\nu}$ and the field strength tensor $F_{M\,\mu\nu}$ are defined as usual for the electromagnetic field
\begin{align}
    \nonumber {T_{M}}^{\, \mu}{}_{\nu}\equiv &\,  g^{\mu\alpha}{T_{M}}_{\,\alpha\nu}=\frac{1}{4\,\pi} \left({F_{M}}^{\,\mu\alpha}F_{M\, \nu\alpha}-\frac{1}{4}F_{M\,\alpha\beta}F_{M}{}^{\alpha\beta}\,\delta^{\mu}{}_{\nu} \right)\, , \quad T_{M}\equiv {T_{M}}^{\, \mu}{}_{\mu}=0\, ,  \\ 
    \nonumber F_{M\,\mu\nu}\equiv &\,  \overline{\nabla}_{\mu} A_{M\, \nu}-\overline{\nabla}_{\nu}A_{M\, \mu}\, . 
\end{align}
Only the metric $g_{\mu\nu}$ is coupled to matter and $f_{\mu\nu}$ is a Kerr-(A)dS solution. The trace-reversed bigravity equations take the form (\ref{trace_rev_matter}), where we have an energy-momentum tensor that comes from the coupling of $g_{\mu\nu}$ to the electromagnetic field. If we ignore the tensors $T^{\mu}{}_{\nu}$ and $\mathcal{T}^{\mu}{}_{\nu}$ that source the solutions focusing only on the non-localized part of the equations, we have
\begin{align}
    \nonumber & R^{\mu}{}_{\nu} - \frac{{\kappa_{g}}^{2}}{2}{\check{T}_{M}{}}^{\,\mu}{}_{\nu}= \Lambda \, \delta^{\mu}{}_{\nu} = Q^{\mu}{}_{\nu} - \frac{1}{2}Q \, \delta^{\mu}{}_{\nu} \, , \quad {T_{M}}^{\,\mu}{}_{\nu} \ne 0 \, , \\
    \nonumber
    & \mathcal{R}^{\mu}{}_{\nu} = \frac{\Lambda}{C^{2}}\delta^{\mu}{}_{\nu} = \mathcal{Q}^{\mu}{}_{\nu} - \frac{1}{2}\mathcal{Q}\, \delta^{\mu}{}_{\nu},
\end{align}
which holds when we use (\ref{knads_sol}). For $f_{\mu\nu}$ we get the same behaviour as in the Kerr-(A)dS as we expected, where the Ricci and the interaction parts are related by the cosmological constant. For the other metric, since ${T_{M}}^{\,\mu}{}_{\nu} \ne 0$, we observe that the Ricci and the interaction part compensate the electromagnetic field contribution which couples to $g_{\mu\nu}$. The double copy equations are
\begin{align}
    \nonumber  &\,^{(g)}\overline{\mathcal{E}}(h^{\mu}{}_{\nu}) - h^{\mu\lambda} \overline{R}_{\lambda \nu}  = \, \frac{\kappa_{g}}{2}{\check{T}_{M}{}}^{\mu}{}_{\nu}\, , \quad  && h_{\mu\nu}=\frac{\kappa_{g}}{2}\frac{1}{8\,\pi\,\Sigma}\left(2m_{1}r-q^{2}\right)k_{\mu}k_{\nu}\, ,  \\
    \nonumber &\,^{(f)}\overline{\mathcal{E}}(\mathscr{h}^{\mu}{}_{\nu}) - \mathscr{h}^{\mu\lambda} \overline{\mathcal{R}}_{\lambda\nu} =\,  0 \, , \quad && \mathscr{h}_{\mu\nu}=\frac{\kappa_{f}}{2}\frac{m_{2}r}{4\,\pi\,\Sigma}k_{\mu}k_{\nu} \, , 
\end{align}
with corresponding single copy equations given by
\begin{align}
    \nonumber  \overline{\nabla}_{\lambda} F^{\lambda\mu} & \, =  {J_{A}}^{\mu}+g_{g}\, J_{M}{}^{\,\mu}\equiv {J_{S}}^{\mu} \,,  \quad && A_{\mu}=g_{g}\frac{1}{8\,\pi\,\Sigma}\left(2m_{1}r-q^{2}\right)k_{\mu} \, ,\\
    \nonumber \overline{\nabla}_{\lambda} \mathcal{F}^{\lambda\mu} & \, =0 \, , \quad &&\mathcal{A}_{\mu}=g_{f}\frac{m_{2}r}{4\,\pi\,\Sigma}k_{\mu} \, , 
\end{align}
where we have defined:
\begin{align}
    \nonumber & {J_{A}}^{\mu} \equiv -\left[\frac{\overline{R}}{6}A^{\mu} + \frac{1}{\alpha}K^{\nu}\left(X^{\mu}{}_{\nu}+Y^{\mu}{}_{\nu}\right)\right]\, , \quad 
    {J_{M}}^{\, \mu} \equiv \frac{-2}{\alpha}K^{\nu}\,{\check{T}_{M}{}}^{\mu}{}_{\nu}\, ,\\
    \nonumber & {J_{S}}^{\mu} \equiv {J_{A}}^{\mu}+\frac{\kappa_{g}}{2}J_{M}{}^{\,\mu} = -\frac{2}{\Sigma^{2}}\,\phi_{M}\, \Bigg(2\left(r^{2}+a^{2}\right)-\Sigma,0,0,a\, (\Delta+\xi)\Bigg)\, .
\end{align}
For the zeroth copy in this case we have that
\begin{align}
    \nonumber \overline{\nabla}^{2}\phi_{g}-\frac{\overline{R}}{6}\phi_{g} & \, = \frac{\overline{R}}{6}\frac{\left(r^{2}+a^{2}\right)}{\Sigma}\phi_{M} + j_{A}+y_{g}\,j_{M} \equiv \, j_{Z} \, , \, &&\phi_{g} =y_{g}\frac{1}{8\,\pi\,\Sigma}\left(2m_{1}r-q^{2}\right)\,, \\  
    \nonumber & \,^{(f)}\overline{\nabla}^{2}\phi_{f}-\frac{\overline{\mathcal{R}}}{6}\phi_{f} \, =  \,0 \, ,  &&\phi_{f} \, =y_{f}\frac{m_{2}r}{4\,\pi\,\Sigma}\, .  
\end{align}
where we have the following definitions
\begin{align}
    \nonumber & j_{Z}\equiv \,\frac{1}{\alpha}K_{\mu}{J_{S}}^{\mu}+\frac{\overline{R}}{6}\frac{\left(r^{2}+a^{2}\right)}{\Sigma}\phi_{M}\, , \quad j_{A}\equiv \frac{1}{\alpha}K_{\mu}{J_{A}}^{\mu}\, , \quad j_{M}\equiv \frac{1}{\alpha}K_{\mu}{J_{M}}^{\mu} \, ,\\
    \nonumber
    &\frac{1}{\alpha^2}K_{\mu}Z^{\mu} \, = \, - \frac{\overline{R}}{6}\left(\phi_{g}+\frac{r^{2}+a^{2}}{\Sigma}\phi_{M}\right). &
\end{align}
We note that neither of the restrictions (\ref{stat_constraints}) nor (\ref{sols_dc_1}) hold for this solution. For the double copy equations we get (\ref{stat_eqs}) modified with the source term which comes from the electromagnetic energy-momentum tensor. For the spin-1 and spin-0 fields we arrive to the Maxwell and conformally coupled equations with source terms which affect non-trivially to the equations. These extra terms that arise when coupling matter to the solutions were discussed in \cite{Bah:2019sda,Alkac:2021bav}. In the context of general relativity, these solutions have not been proven to come from a double copy.  We can observe how these sources complicate the process of obtaining this charged solution from a double copy procedure.

The Reissner-Nordstr\"om solution in (A)dS background is obtained when we consider the non-rotating case, $a=0$. Then, the resulting sources are
\begin{align}
    {J_{A}}^{\mu}=0 \, , \quad \ \ \ \ \ {J_{S}}^{\mu}=\frac{\kappa_{g}}{2}J_{M}{}^{\,\mu}=\frac{2}{r^{2}}\phi_{M}\left(-1,0,0,0 \right) \,, \quad j_{M} = \frac{2}{r^{2}}\, \phi_{M},
\end{align}
which implies that the single and zeroth copy equations for the metric $g_{\mu\nu}$ reduce to
\begin{align}
     \nonumber \overline{\nabla}_{\lambda} F^{\lambda\mu} \, =  g_{g}\, J_{M}{}^{\,\mu} \,,  \qquad \ \ \ \ \ \overline{\nabla}^{2}\phi_{g}-\frac{\overline{R}}{6}\, \phi_{g} \, = y_{g}\,j_{M} \, .
\end{align}
In this case, it is worth noting that the single copy field $A_{\mu}$ is coupled to the source $J_{M}{}^{\,\mu}$, and for the zeroth copy, the conformally coupled scalar field $\phi_{g}$ is coupled to the source $j_{M}$. It is interesting to note that $\phi_{M}$ satisfies the following equation:
$$
\overline{\nabla}^{2}\phi_{M}-\frac{\overline{R}}{6}\phi_{M} - \frac{2}{r^{2}}\phi_{M} \, = 0 \, .
$$

\subsection{Time-dependent solutions}
The time-dependent solutions we present in this work are wave solutions: Siklos-AdS waves and $pp$-waves in bigravity. For the case of general relativity, these were studied in \cite{Carrillo-Gonzalez:2017iyj} within the framework of the double copy formalism. For these time-dependent solutions in bigravity we have that
\begin{align}
\label{waves_constraints}
    X^{\mu}{}_{\nu} = \, Y^{\mu}{}_{\nu} = 0 \, , \quad \mathcal{X}^{\mu}{}_{\nu} = \mathcal{Y}^{\mu}{}_{\nu} = 0 \, , \quad B_{1} = \frac{m^{2} \kappa_{g}}{\kappa^{2}}(C P_{0}) \, , \quad  \mathcal{B}_{1} = \frac{m^{2} \kappa_{f}}{\kappa^{2}}\left(\frac{P_{0}}{C^{5}}\right) \,,
\end{align}
where $P_{0}$ may be nonzero. In this case we will have linear contributions to the interaction terms that depend on the perturbations. When the restrictions in (\ref{waves_constraints}) hold they lead us to the following double, single and zeroth copy equations for the ansatz:
\begin{align}
\label{eom_dszc_2}
    \nonumber \,^{(g)}\overline{\mathcal{E}}(h^{\mu}{}_{\nu}) - \frac{\overline{R}}{4} \, h^{\mu}{}_{\nu} \,  = &\,  - \nonumber B_{1} \left(\kappa_{g}h^{\mu}{}_{\nu}-C^{2}\kappa_{f}\mathscr{h}^{\mu}{}_{\nu}\right) \, , \\
    \nonumber \,^{(f)}\overline{\mathcal{E}}(\mathscr{h}^{\mu}{}_{\nu}) -\,\frac{\overline{\mathcal{R}}}{4} \, \mathscr{h}^{\mu}{}_{\nu} \,\, = & \, \mathcal{B}_{1}\left(\kappa_{g}h^{\mu}{}_{\nu}-C^{2}\kappa_{f}\mathscr{h}^{\mu}{}_{\nu}\right)\, ,\\
    \nonumber \overline{\nabla}_{\lambda}F^{\lambda\mu}+\frac{\overline{R}}{6}A^{\mu} \, = & \, 2 B_{1} \left(\kappa_{g}A^{\mu}-C^{2}\kappa_{f}\mathcal{A}^{\mu}\right) \, , \\
    \nonumber \,^{(f)}\overline{\nabla}_{\lambda}\mathcal{F}^{\lambda\mu} +\frac{\overline{\mathcal{R}}}{6}\mathcal{A}^{\mu} \, = & \, -2\, \mathcal{B}_{1} \left(\kappa_{g}A^{\mu}-C^{2}\kappa_{f}\mathcal{A}^{\mu}\right) \, ,\\
    \nonumber \overline{\nabla}^{2}\phi_{g}\,  = \, &  2 B_{1} \left(\kappa_{g}\phi_{g}-\kappa_{f}\phi_{f}\right) \, , \\
    \,^{(f)}\overline{\nabla}^{2}\phi_{f} \, = \, & - 2\,\mathcal{B}_{1} \left(\kappa_{g}\phi_{g}-\kappa_{f}\phi_{f}\right) \, .
\end{align}
The equations we have in (\ref{eom_dszc_2}) are written in such way that they separate the equations that come from the interaction term between the metrics $g_{\mu\nu}$ and $f_{\mu\nu}$.  If we rewrite the equations to split the contributions of the fields that correspond to each metric we obtain:
\begin{align}
\label{eom_dszc_3}
    \nonumber \,^{(g)}\overline{\mathcal{E}}(h^{\mu}{}_{\nu}) + \frac{1}{2}{{m_{D}}^{2}} \, h^{\mu}{}_{\nu} \, = & \,  \left(C^{2}\,\kappa_{f}\,B_{1}\right)\mathscr{h}^{\mu}{}_{\nu} \, , \qquad &&{m_{D}}^{2} \equiv 2\, \kappa_{g}\, B_{1}-\frac{\overline{R}}{2}, \\
    \nonumber \,^{(f)}\overline{\mathcal{E}}(\mathscr{h}^{\mu}{}_{\nu})  +\,\frac{1}{2}{\mathscr{m}_{D}}^{2} \, \mathscr{h}^{\mu}{}_{\nu} \, = & \, 
     \left(\kappa_{g}\, \mathcal{B}_{1}\right) h^{\mu}{}_{\nu}\, , \qquad && {\mathscr{m}_{D}}^{2} \equiv 2\,C^{2}\,\kappa_{f}\,\mathcal{B}_{1}-\frac{\overline{\mathcal{R}}}{2} \, , \\
    \nonumber \overline{\nabla}_{\lambda}F^{\lambda\mu} -{m_{S}}^{2}A^{\mu} \, = & \, \left(- 4\, C^{2}\,g_{f}\, B_{1}\right) \mathcal{A}^{\mu} \, , \qquad  &&{m_{S}}^{2} \equiv 4\,g_{g}\, B_{1}-\frac{\overline{R}}{6} \, ,\\
    \nonumber \overline{\nabla}_{\lambda}\mathcal{F}^{\lambda\mu}-{\mathscr{m}_{S}}^{2}\mathcal{A}^{\mu} \, = & \,  \left(-4\,g_{g}\,\mathcal{B}_{1}\right) \, A^{\mu} \, , \qquad &&{\mathscr{m}_{S}}^{2} \equiv 4\, C^{2}\,g_{f}\,\mathcal{B}_{1}-\frac{\overline{\mathcal{R}}}{6} \, ,\\     \nonumber \overline{\nabla}^{2}\phi_{g}-{m_{Z}}^{2}\phi_{g}\, = &\, \left(-4\,y_{f}\,B_{1}\right)  \phi_{f} \, , \qquad &&{m_{Z}^{2}} \equiv 4\,y_{g}\, B_{1} \, , \\ 
    \,^{(f)}\overline{\nabla}^{2}\phi_{f}-{\mathscr{m}_{Z}}^{2}\phi_{f}\, = &\, \left(- 4 \,C^{2} \, y_{g}\, \mathcal{B}_{1}\right)  \phi_{g} \, , \qquad &&{\mathscr{m}_{Z}}^{2} \equiv 4\,C^{2}\, y_{f}\,\mathcal{B}_{1} \, ,
\end{align}
so that for the double copy equations we have two interacting spin-2 fields, for the single copy equations we have two interacting spin-1 fields and the zeroth copy equations result to be two interacting spin-0 fields. We have defined the parameters $m^{2}$ and $\mathscr{m}^{2}$ for the equations obtained from the metric $g_{\mu\nu}$ and $f_{\mu\nu}$ respectively. The negative signs accompanying the mass terms in the single and zeroth copy are reversed compared to the usual positive sign, which is due to our choice of signature $(-,+,+,+)$. We may have problems interpreting these quantities as the square of some masses because they are not restricted to be positive. For the double and single copy, these parameters depend on the constant $P_{0}$ (which in turn, depends on the interaction constants $b_{1}, b_{2}$ and $b_{3}$) and on the scalar curvature of the background spacetimes and in zeroth copy they only depend on the interaction constants, not on the scalar curvatures. The fields we obtained from the double copy procedure have a term proportional to this ``mass squared'' parameters. The constants on the right hand side of these equations, $B_{1}$ and $\mathcal{B}_{1}$ depend on $P_{0}$. 

Thus we notice that for the time-dependent solutions the single copy equations contain terms proportional to the scalar curvature of the background metrics and the zeroth copy does not have this term. These results are different from the stationary case where the zeroth copy are the equations which contain a term proportional to the background curvature, whereas the single copy equations do not contain such term. These results coincide with the ones obtained in \cite{Carrillo-Gonzalez:2017iyj} for these type of solutions in general relativity.

\subsubsection{Decoupling the system of equations for time-dependent solutions}

If we perturb the metrics in bigravity around curved background metrics, the perturbations result to be a superposition of massive and the massless fields spin-2 fields \cite{deRham:2014zqa}. Because in the stationary case we note that the equations of motion are already decoupled, we are interested in decoupling the systems of differential equations we have obtained for the time-dependent solutions in order to obtain the massive and massless fields. Our purpose is to decouple the equations using a redefinition of the fields and to extract the massive and massless parts at each level in the double copy.

Before proceeding with the decoupling, we observe that we are using distinct derivative operators with index up in the equations, which comes from the fact that we raised indices using different metrics. If we are interested in decoupling the equations just by a redefinition of the fields but preserving the structure of the differential equations, to have the same differential operators acting on the fields, which will permit us factorize the fields and to decouple them. Given the ansatz in (\ref{type_b_ansatz_2}), the operators we use are related by a proportionality constant, for which we can express one operator in terms of the other. After rewriting the equations of motion using only one covariant derivative operator we can redefine the fields as:
\begin{align}
\label{redef_ads_1}
    \nonumber & h^{+\,\mu}{}_{\nu}\equiv -\frac{\kappa_{g}\kappa_{f}}{\kappa^{2}}\left(\kappa_{f}\,h^{\mu}{}_{\nu}+C^{2}\,\kappa_{g}\,\widetilde{\mathscr{h}}^{\mu}{}_{\nu}\right)\, , \quad h^{-\,\mu}{}_{\nu}\equiv \kappa_{g}\, h^{\mu}{}_{\nu}-\kappa_{f}\,\widetilde{\mathscr{h}}^{\mu}{}_{\nu}\, ,\\
    \nonumber & A^{+\, \mu} \equiv -\frac{\kappa_{g}\kappa_{f}}{\kappa^{2}}\left(\kappa_{f}\, A^{\mu}+C^{2}\, \kappa_{g}\,\widetilde{\mathcal{A}}^{\mu}\right) \, , \quad  A^{-\, \mu} \equiv \kappa_{g}\,A^{\mu}- \kappa_{f}\,\widetilde{\mathcal{A}}^{\mu} \, ,\\
    &\phi^{+} \equiv -\frac{\kappa_{g}\kappa_{f}}{\kappa^{2}}\left(\kappa_{f} \, \phi_{g}+C^{2}\, \kappa_{g} \, \phi_{f}\right)\, , \quad  \phi^{-} \equiv \kappa_{g}\,\phi_{g}-\kappa_{f}\,  \phi_{f}\, ,
\end{align}
where we have used the following definitions:
\begin{align}
    \nonumber \widetilde{\mathscr{h}}^{\mu}{}_{\nu}\equiv \overline{g}^{\mu\alpha}\mathscr{h}_{\alpha\nu}= C^{2}\mathscr{h}^{\mu}{}_{\nu} \, , \quad  \widetilde{\mathcal{A}}^{\mu}\equiv \overline{g}^{\mu\alpha}\mathcal{A}_{\alpha}= C^{2}\mathcal{A}^{\mu} 
\end{align}
instead of $\mathscr{h}^{\mu}{}_{\nu}$ and $\mathcal{A}^{\mu}$. With these redefinitions of the fields, the equations for the double, single and zeroth copies can be written as:
\begin{align}
\label{decoup_eqns}
    \nonumber  \,^{(g)}\overline{\mathcal{E}}(h^{+\mu}{}_{\nu})\,  +\frac{1}{2}{m^{+}_{D}}^{2} h^{+\,\mu}{}_{\nu} \, = & \, 0  \, , \qquad {m^{+}_{D}}^{2} \equiv - \frac{\overline{R}}{2} \, , \\
    \nonumber \,^{(g)}\overline{\mathcal{E}}(h^{-\mu}{}_{\nu}) + \frac{1}{2} {m^{-}_{D}}^{2} \, h^{-\, \mu}{}_{\nu} \, = & \, 0  \, , \qquad {m^{-}_{D}}^{2} \equiv \widehat{m}^{2} - \frac{\overline{R}}{2} \, , \\
    \nonumber \overline{\nabla}_{\lambda}F^{+\,\lambda\mu} - {m^{+}_{S}}^{2} \, A^{+ \, \mu} = &\,  0 \, , \qquad {m^{+}_{S}}^{2} \equiv -\frac{\overline{R}}{6} \, ,\\
    \nonumber \overline{\nabla}_{\lambda}F^{-\,\lambda\mu} - {m^{-}_{S}}^{2}\,  A^{- \, \mu} \, = & \, 0 \, , \qquad {m^{-}_{S}}^{2} \equiv \widehat{m}^{2} -\frac{\overline{R}}{6}\, , \\
    \nonumber \overline{\nabla}^{2}\phi^{+}= &\, 0 \, , \qquad {m^{+}_{Z}}^{2} \equiv 0  \, , \\ 
    \overline{\nabla}^{2}\phi^{-} - {m^{-}_{Z}}^{2} \, \phi^{-} \, = &\, 0 \, , \qquad {m^{-}_{Z}}^{2} \equiv \widehat{m}^{2} \, ,
\end{align}
where we have used $\overline{R}=C^{2}\overline{\mathcal{R}}$ and the mass $\widehat{m}^{2}$ which is the mass of the zeroth copy
\begin{align} 
    \nonumber {m^{-}_{Z}}^{2} \equiv 2 \left(\kappa_{g} \,B_{1} + C^{4}\,\kappa_{f}\,\mathcal{B}_{1}\right)= \frac{2m^{2}P_{0}}{C\kappa^{2}}(C^{2} {\kappa_{g}}^{2}+{\kappa_{f}}^{2})\, = \widehat{m}^{2} , 
\end{align}
and it is given in terms of the Fierz-Pauli mass. This exact decoupling of the fields for time-dependent solutions in bigravity, employing the ansatz (\ref{type_b_ansatz_2}), was first presented in \cite{Ayon-Beato:2018hxz}. The decoupled equations in (\ref{decoup_eqns}) stem from this result, reformulated to allow an interpretation within the framework of the double copy. The factor of two appearing in the mass $\widehat{m}^{2}$ compared to the reference is a consequence of our definition choice for $P_{0}$.

In (\ref{decoup_eqns}), we have the decoupled equations for the decoupled fields (\ref{redef_ads_1}): At the level of the double and single copy equations we obtain that both ``$+$'' and ``$-$'' fields obey differential equations of massive fields. The mass of these ``$-$'' fields depends on the background curvature and the interaction constants and for the ``$+$'' fields it depends only on the scalar curvature of the background. For the zeroth copy we have that the dependence of the mass on the curvature disappears and we have a massless scalar field $\phi^{+}$ and a massive scalar field $\phi^{-}$ with mass proportional to the square of the Fierz-Pauli mass and the gravitational constants. 

An interpretation of the fields in the equation of bigravity as particles would be more straightforward in flat backgrounds ($\Lambda_{g}= \Lambda_{f} = 0$), in whose case we would obtain equations consisting in one massive and one massless field at each level. In such case it becomes easier to elucidate the massive degrees of freedom. We have two decoupled spin-2 equations, one for a massive field $h^{-\mu}{}_{\nu}$ and one for the massless field $h^{+,\mu}{}_{\nu}$ in the double copy equations and $5+2=7$ degrees of freedom corresponding to the massive and massless spin-2 fields respectively, which is consistent with the fact that we have $5+2$ propagating modes in bigravity. For the single copy fields we have $3+2=5$ degrees of freedom related to the massive $A^{-\,\mu}$ and massless $A^{+\,\mu}$ spin-1 fields. For the zeroth copy we have $1+1=2$ degrees of freedom corresponding to the massive and massless scalar fields. We proceed to present a couple of time-dependent solutions in bigravity.

\subsubsection{Example: Separable Siklos-AdS waves in bigravity}

Gravitational waves in bigravity in flat spacetimes were studied in \cite{Mohseni:2012ug}. In the present article we will study another example of time-dependent solutions, the AdS waves, which are exact gravitational waves propagating on an AdS background. This comes from the fact that the profiles $\phi$ in the ansatz (\ref{type_b_ansatz_2}) satisfy wave equations on AdS space. The case of AdS waves in bigravity was studied in \cite{Ayon-Beato:2018hxz} using Poincar\'e coordinates. The line element for AdS waves uses AdS spacetime as background and a null vector, such that in these coordinates we have explicitly:
\begin{align}
    \nonumber d\overline{s}^2=\frac{l^2}{y^2}\left(-2dudv+dx^2+dy^2\right) \, , \quad k_{\mu}dx^{\mu}=-\frac{l}{y}du \, .
\end{align}
In order to construct the full metric of the AdS waves for both metrics, we also need to specify the scalar profiles $\phi_{g}$ and $\phi_{f}$. 

In \cite{Ayon-Beato:2018hxz}, the authors presented a general solution for the profiles, but for the purposes of this work, we will focus on a family of separable solutions they studied, which is the case where the profiles $\phi_{g}$ and $\phi_{f}$ can be decomposed as a sum of two separable solutions of the wavefront coordinates. This permits to decouple the excitations and to analyse the physical propagating modes of the theory. For these separable solutions, we have two cases: in one case the profiles lead to one massive and one massless excitation, called ``massive profiles'', and the ``massless profiles'' which result in two massless excitations. In the latter case, we refer to the fact that, even if the flat-space graviton mass $m$ is nonzero, we can have zero mass modes by imposing the interaction constant $P_{0}$ to be zero. For the massive profiles, by a redefinition of the fields the differential equations satisfied by $\phi_{g}$ and $\phi_{f}$ can be decoupled into two scalar fields: one which is an exact massless excitation and satisfies a Klein-Gordon equation on AdS and one exact massive excitation which satisfies the massive Klein-Gordon equation on AdS with a specific mass expressed in terms of the mass $m$, also known as the Fierz-Pauli mass $m$. We will use the functions $F_{1}(u,x,y)$ and $F_{2}(u,x,y)$ as defined in \cite{Ayon-Beato:2018hxz} so that for the ansatz one has $\phi_{g}=-\left({1}/{\kappa_{g}}\right)F_{1}(u,x,y)$ and $\phi_{f}=-\left({1}/{\kappa_{f}}\right)F_{2}(u,x,y)$, where we also have to take into account the gravitational couplings in $F_{1}$ and $F_{2}$.

\vskip .5truecm

\begin{center}
    \textbf{Massless profiles ${m^{-}_{Z}} = 0$ ($P_{0} = 0$)}
\end{center}
The solutions found in \cite{Ayon-Beato:2018hxz} for massless profiles for $F_{1}(u,x,y)$ and $F_{2}(u,x,y)$ give us the following form for $\phi_{g}$ and $\phi_{f}$:
\begin{align}
    \nonumber &\phi_{g}(u,x,y) =-\left(\frac{\kappa_{g}}{2}\right) \frac{2}{{\kappa_{g}}^{2}}f_{3}(u)\left(\frac{y}{l}\right)^{3},  \\
    \nonumber &\mathsf{\phi}_{f}(u,x,y) =-\left(\frac{\kappa_{f}}{2}\right) \frac{2}{{\kappa_{f}}^{2}}\left[h_{3}(u)\left(\frac{y}{l}\right)^{3}+\frac{h_{2}(u)}{l^{2}}\left(x^{2}+y^{2}\right)+h_{1}(u)\frac{x}{l}+h_{0}(u) \right]\, ,
\end{align}
where  $h_{0}, h_{1},h_{2}, h_{3}$ and $f_{3}$ are arbitrary functions. For this solution we have that:
\begin{align}
    \nonumber & P_{0}=0, 
    \quad P_{1}=-\frac{\kappa}{m^{2}\kappa_{g}} \Lambda_{g} \, , 
    \quad P_{2}=-\frac{C^{3} \kappa}{m^{2}\kappa_{f}}\Lambda_{f} \, , \quad \Lambda_{g} = -\frac{3}{l^{2}}= C^{2} \Lambda_{f} \, , \\
    \nonumber \Rightarrow  \ \ \ \ \quad &  B_{0}= \frac{-\Lambda_{g}}{\kappa_{g}} =\frac{C^{4}\,\kappa_{f}}{\kappa_{g}} \mathcal{B}_{0} \, , \quad B_{1}=  \mathcal{B}_{1} = 0 \, .
\end{align}
These massless profiles of the Siklos-AdS solution in bigravity satisfy the equations in (\ref{eom_dszc_3}) with $P_{0}=0$ so that the fields are decoupled at each level of the double copy. Here we use the null Killing vector of the solution to obtaining the single and zeroth copy equations. Then, with these profiles, the null vector $k_{\mu}$ and the rules in (\ref{sing_replace}) and (\ref{zero_replace}),  we can construct the corresponding double, single and zeroth copy fields. In the single copy equations we have a term that is proportional to the scalar curvature of each metric which breaks the gauge invariance. At the zeroth copy level, we have that each of the functions $\phi_{g}$ and $\phi_{f}$ satisfies the massless Klein-Gordon equation. We do not have a term proportional to the background curvature in the zeroth copy equations which we had for the black hole solutions.

\vskip .5truecm

\begin{center}
    \textbf{Massive profiles ${m^{-}_{Z}}
    \ne 0$ ($P_{0}\ne 0$)}
\end{center}
The expression for the massive profiles $\phi_{g}$ and $\phi_{f}$ obtained in \cite{Ayon-Beato:2018hxz} are linear combinations of the decoupled solutions of the form $\phi^{\pm}=X_{\pm}(u,x)+Y_{\pm}(u,y)$. The resulting potentials in the Kerr-Schild ansatz are
\begin{align}
    \nonumber &\phi_{g}=-\left(\frac{\kappa_{g}}{2}\right)\frac{2}{{\kappa_{f}}^2+C^{2}{\kappa_{g}}^{2}}\bigg[\frac{\kappa^{2}}{{\kappa_{g}}^{2}} f_{3}(u)\left(\frac{y}{l}\right)^{3} -\,C^{2} \left(h_{1}(u)\left(\frac{y}{l}\right)^{\,\rho_{+}}+h_{2}(u)\left(\frac{y}{l}\right)^{\,\rho_{-}}\right)\bigg], \\
    \nonumber &\phi_{f}=-\left(\frac{\kappa_{f}}{2}\right)\frac{2}{{\kappa_{f}}^{2}+C^{2}{\kappa_{g}}^{2}}\bigg[\frac{\kappa^{2}}{{\kappa_{f}}^{2}} f_{3}(u)\left(\frac{y}{l}\right)^{3} +\left(h_{1}(u)\left(\frac{y}{l}\right)^{\,\rho_{+}}+h_{2}(u)\left(\frac{y}{l}\right)^{\,\rho_{-}}\right)\bigg] \, ,
\end{align}
where $h_{1},h_{2}$ and $f_{3}$ are arbitrary functions. Moreover here, 
\begin{align}
    \nonumber \rho_{\pm}\equiv\frac{3}{2}\pm l \sqrt{\widehat{m}^{2}-{m_{BF}}^{2}}\, , \quad {m_{BF}}^{2}\equiv -\frac{9}{4l^{2}} \, , 
\end{align}
where $\rho_{\pm}$ are the roots of the characteristic polynomial determining the linearly independent
solutions of the ordinary Euler equation for the massive modes and we have defined the Breitenlohner-Freedman bound ${m_{BF}}^{2}$ which gives the lowest possible value for the square of the mass that a stable scalar field can take on a AdS background \cite{Breitenlohner:1982jf}. Using these massive profiles and the null vector $k_{\mu}$ we can construct the double and single copy fields. We obtain that, given arbitrary functions, for this solution we have that
{\small\begin{align}
    \nonumber &P_{0}=\frac{C \,\kappa}{2 m^{2} \left(C^{2} {\kappa_{g}}^{2}+{\kappa_{f}}^{2} \right)} \, \widehat{m}^{2} \, , 
    \quad P_{1}=-\left(-\frac{3}{l^{2}}   \right)\frac{\kappa^{2}}{m^{2}{\kappa_{g}}^{2}} \, , 
    \quad P_{2}=-\left(-\frac{3}{l^{2}}   \right)\frac{C \kappa^{2}}{m^{2}{\kappa_{f}}^{2}} \, , \\
    \nonumber \quad & B_{0}= \frac{-\Lambda_{g}}{\kappa_{g}} =\frac{C^{4}\,\kappa_{f}}{\kappa_{g}} \mathcal{B}_{0}  \, ,  \quad  B_{1}=\frac{ C^{2}\,\kappa_{g}}{2\,\left(C^{2}\, {\kappa_{g}}^{2}+{\kappa_{f}}^{2}\right)}\, \widehat{m}^{2} =\frac{C^{6}\kappa_{g}}{\kappa_{f}} \mathcal{B}_{1} \,, \quad \Lambda_{g} = -\frac{3}{l^{2}}= C^{2} \Lambda_{f} \, ,
\end{align}}and the double, single and zeroth copy equations satisfy the equations in (\ref{eom_dszc_3}) respectively, and also the decoupled equations in (\ref{decoup_eqns}). Here we are using a null Killing vector of the solution to obtain the single and zeroth copy equations. Using the redefinitions in (\ref{redef_ads_1}) to decouple massive profiles:
\begin{align}
    \nonumber \phi^{+}(u,x,y)= f_{3}(u)\left(\frac{y}{l}\right)^{3} \, ,  \quad \phi^{-}(u,x,y) =h_{1}(u)\left(\frac{y}{l}\right)^{\,\rho_{+}}+h_{2}(u)\left(\frac{y}{l}\right)^{\,\rho_{-}} \, ,
\end{align}
where the masses of the zeroth copy, i.e. the masses of the scalar profiles $\phi^{+}$ and $\phi^{-}$ are:
\begin{align} 
    \nonumber  {m^{+}_{Z}}^{2} = 0\, , \quad   {m^{-}_{Z}}^{2} = \widehat{m}^{2}\, , 
\end{align}
and for the single and double copy the corresponding masses are shown in (\ref{decoup_eqns}). If we saturate the Breitenlohner-Freedman bound, i.e., if we have $\widehat{m}^{2}={m_{BF}}^{2}$, we obtain a logarithmic behavior due to the multiplicity of the roots $\rho_{\pm}$. In such case we have
\begin{align}
    \nonumber \phi^{+}(u,x,y)=\, f_{3}(u)\left(\frac{y}{l}\right)^{3} \, ,  \quad \phi^{-}(u,x,y) =\left(\frac{y}{l}\right)^{\frac{3}{2}}\left(h_{1}(u)+h_{2}(u)\ln\frac{y}{l}\right) \, ,
\end{align}
and the equations in (\ref{decoup_eqns}) hold with: 
\begin{align} 
    \nonumber  {m^{+}_{Z}}^{2} = 0\, , \quad   {m^{-}_{Z}}^{2} = {m_{BF}}^{2} = -\frac{9}{4l^{2}} \, . 
\end{align}
For this particular case, the masses corresponding to the single and double copy are given by:
\begin{align} 
    \nonumber {m^{+}_{D}}^{2} = \frac{6}{l^{2}}=&\,-2\Lambda_{g} \, , \quad   {m^{-}_{D}}^{2} = \frac{15}{4l^{2}} = -\frac{5}{4}\Lambda_{g} \, ,\\ 
    \nonumber {m^{+}_{S}}^{2} = \frac{2}{l^{2}} =& \, -\frac{2}{3}\Lambda_{g}\, , \quad   {m^{-}_{S}}^{2}= -\frac{1}{4l^{2}} = \frac{1}{12}\Lambda_{g} \, . 
\end{align}
As stated before, with this example we note that in curved spacetime, the background scalar curvature contributes to the masses defined for the double and single copy and does not appear in the masses of the zeroth copy.

\subsubsection{Example: \textit{pp}-waves in bigravity}
For general relativity, in the context of the double copy, $pp$-waves were studied in \cite{Carrillo-Gonzalez:2017iyj}. Also, in \cite{Gurses:2018ckx}, the authors worked with the coupled system of the Einstein-Yang-Mills-Maxwell and discussed the possibility to extended the formalism to the Kerr-Schild-Kundt class, in which case, it is shown that a null fluid is needed for that case for the correspondence to work. In order to explore the double copy equations for these solutions in bigravity, in this work we will focus on the approach to $pp$-waves as outlined in \cite{Carrillo-Gonzalez:2017iyj}, while also extending the analysis to incorporate the solutions derived for $pp$-waves in bigravity, as documented in \cite{Ayon-Beato:2018hxz}.

We can consider waves propagating in a flat background, i.e., with vanishing cosmological constant $\Lambda_{g}=0=\Lambda_{f}$ . Here we use:
\begin{align}
    \nonumber d\overline{s}^2=\left(-2dudv+dx^2+dy^2\right) \, , \quad k_{\mu}dx^{\mu}=-du \, .
\end{align}
In \cite{Ayon-Beato:2018hxz}, there are presented solutions for $\widehat{m}\ne 0$, so that in this case we will only consider massive profiles. The massive profiles we will use for $pp$-waves are
{\small\begin{align}
    \nonumber &\phi_{g}(u,{\vec{x}}) =-\left(\frac{\kappa_{g}}{2}\right)\frac{2}{C^{2} {\kappa_{g}}^{2}+{\kappa_{f}}^{2}}\left[\frac{\kappa^{2}}{{\kappa_{g}}^{2}} f_{2}(u)(x^{2}-y^{2}) -C^{2}\left(\vec{h}_{+}(u)\cdot e^{\widehat{m}\vec{x}}+\vec{h}_{-}(u)\cdot e^{-\widehat{m}\vec{x}}\right)\right] \, , \\
    \nonumber &\phi_{f}(u,{\vec{x}}) =-\left(\frac{\kappa_{f}}{2}\right)\frac{1}{C^{2} {\kappa_{g}}^{2}+{\kappa_{f}}^{2}}\left[\frac{\kappa^{2}}{{\kappa_{f}}^{2}} f_{2}(u)(x^{2}-y^{2}) +\left(\vec{h}_{+}(u)\cdot e^{\widehat{m}\vec{x}}+\vec{h}_{-}(u)\cdot e^{-\widehat{m}\vec{x}}\right)\right] \, , 
\end{align}}where $e^{\pm\widehat{m}\vec{x}}=(e^{\pm\widehat{m}x},e^{\pm\widehat{m}y})$. The functions of the retarded time are arbitrary and the parameter $\widehat{m}^{2}$ are as presented in \cite{Ayon-Beato:2018hxz}. The constants for this family of solutions are given by
\begin{align}
    \nonumber & P_{0}=\frac{C \,\kappa^{2}}{2 m^{2} \left(C^{2} {\kappa_{g}}^{2}+{\kappa_{f}}^{2} \right)} \, \widehat{m}^{2}\, , 
    \quad P_{1}= P_{2}=0 \quad (\Rightarrow B_{0}=\mathcal{B}_{0}=0 )  \, .
\end{align}
The massless and massive fields are:
\begin{align}
    \nonumber \phi^{+}(u,\vec{x})=f_2(u)(x^2-y^2)\, ,\quad  \phi^{-}(u,\vec{x})=\vec{h}_+(u)\cdot e^{\widehat{m}\vec{x}}+\vec{h}_-(u)\cdot e^{-\widehat{m}\vec{x}}\,
\end{align}
and the parameter $\widehat{m}^{2}$ is nonzero and related to the masses for the decoupled fields as:
\begin{align} 
    \nonumber  {m^{+}_{Z}}^{2} = &\,  0\, , \quad   {m^{-}_{Z}}^{2} = \widehat{m}^{2}\, , \\
    \nonumber {m^{+}_{S}}^{2} = & \, 0 \, , \quad   {m^{-}_{S}}^{2}= \widehat{m}^{2}\, , \\
    \nonumber {m^{+}_{D}}^{2} = &\,0 \, , \quad   {m^{-}_{D}}^{2} = \widehat{m}^{2} \, .
\end{align}
Using the null Killing vector of the solution, the equations (\ref{eom_dszc_3}) and the decoupled equations in (\ref{decoup_eqns}) are satisfied with the above masses. As expected for the flat case, we have one massive and one massless field at each level. As mentioned in \cite{Ayon-Beato:2018hxz}, the massless profile $\phi^{+}$ corresponds to the linearly polarized plane wave contribution of general relativity.  On the other hand, the massive profile $\phi^{-}$ is associated with a combination of Yukawa exponential decays and growths in each wavefront direction, exhibiting the characteristics of standard massive modes in flat spacetime.

\section{Final Remarks}\label{final}
We have presented, in the framework of the double copy, the bigravity equations of motion for the generalized Kerr-Schild ansatz (\ref{type_b_ansatz_2}), which was studied in \cite{Ayon-Beato:2015qtt}, and considers that the background metrics are proportional and that the null vectors of both metrics are the same. This particular ansatz permits to calculate the interaction tensors explicitly in terms of the perturbations in such way that we only have contributions up to linear terms in $h_{\mu\nu}$ and $\mathscr{h}_{\mu\nu}$ in the interaction. The resulting equations can be written as a pair of equations for two spin-2 interacting fields in a curved background spacetime (\ref{type_b_ads_eom_1}), which would correspond to the double copy equations. In order to study the Kerr-Schild classical double copy for bigravity in curved backgrounds we focus on maximally symmetric spacetimes, which permit us rewrite the spin-2 equations as in (\ref{eom_sc_1}) after contracting the equations using Killing vector fields. Such equation depends on the two coupled spin-1 fields $A_{\mu}$ and $\mathcal{A}_{\mu}$ and other tensors, which would be our single copy equations. By the same procedure we obtain the zeroth copy equations (\ref{eom_zc_1}) which depend on the coupled spin-0 fields $\phi_{g}$ and $\phi_{f}$ and other tensors. We have  considered stationary and time-dependent solutions and the equations obtained at the level of the double, single and zeroth copy simplify when considering some examples. These results also applies to flat background spacetimes. 

The cases we study in the stationary solutions are the Schwarzschild-(A)dS and the Kerr-(A)dS solution. For these solutions, the equations we obtain are (\ref{stat_eqs}), where we have a pair of Maxwell equations in the single copy and the zeroth copy equations are two conformally coupled scalar fields. Both fields are decoupled in the stationary case and the interaction tensors do not contain contributions which are linear in the perturbations. We also consider a stationary solution where one of the metrics is coupled to the electromagnetic field: one of the metrics is a Kerr-Newman-(A)dS solution and the other is a Kerr-(A)dS solution. In such case we arrive to single and zeroth copy equations that contain source terms which are not straightforward to interpret in terms of quantities of the fields we have defined. We mention that in the stationary vacuum examples, the interaction between the two metrics results to be a cosmological constant contribution to the equations of motion, where the Ricci part of the trace-reversed equations equals the interaction term which is the cosmological constant term. In general relativity we have to add the cosmological term when considering maximally symmetric spacetimes. For the  solutions to the bigravity equations of motion presented here, we do not have to add this term by hand as it results from the interaction between the metrics. In the Kerr-Newman-(A)dS solution we see that a similar phenomena takes place: the Ricci and electromagnetic contributions to the equations of motion compensate to equalize the interaction between the metrics so that the bigravity equations hold, which for these black holes is the cosmological constant contribution for each particular metric. 

For time-dependent solutions we focus on the Siklos-AdS waves and $pp$-waves. The equations these solutions follow are the ones in (\ref{eom_dszc_3}) and also the decoupled equations in (\ref{decoup_eqns}) with specific interaction constants. In these cases we have linear interactions, which means that we have terms that are linear in the perturbations contained in the interaction tensors. We note that we do not have a term proportional to the background curvature in the zeroth copy level, as we have it at the level of the single copy. These results are consistent with the ones presented in \cite{Carrillo-Gonzalez:2017iyj} for general relativity. By decoupling the differential equations we obtain a couple of Proca fields for the single copy. For the zeroth copy we obtain that the field $\phi^{+}$ is massless and the field $\phi^{-}$ is massive with mass proportional to the Fierz-Pauli mass and which depends on the interaction coefficients $b_{1}, b_{2}$ and $b_{3}$ which suggests that the ``mass'' of these fields is caused by the interaction between the two metrics in bigravity. 

In order to extend these results we need to study more solutions in bigravity. We could consider, for example, more black hole solutions, e.g., a Taub-NUT-like solution as in \cite{Luna:2015paa} for bigravity and develop a formalism using a double Kerr-Schild ansatz, or we could also consider another type of classical double copy, like the Weyl double copy in the context of bigravity.

Also, a formalism with non-linear interactions or with a modified bigravity action as in \cite{Gialamas:2023lxj} may be another possible direction. Nevertheless, even though some aspects of the double copy relations in general relativity have been understood, there is still not a complete understanding of it, so in this work we tried to broaden the domain of applicability of the double copy to the theory of bimetric massive gravity.

\acknowledgments
It is a pleasure to thank E. Ay\'on-Beato, M. Carrillo Gonz\'alez, E. Chac\'on, A. Luna and R. Monteiro for useful feedback and enlightening comments. C. Ramos thanks Conahcyt for scholarship No. 833288.




\clearpage




\end{document}